\documentclass[fleqn,usenatbib]{mnras}

\usepackage[T1]{fontenc}
\usepackage{ae,aecompl}

\usepackage{graphicx}	% Including figure files
\usepackage{amsmath}	% Advanced maths commands
\usepackage{amssymb}	% Extra maths symbols

\title[Molecular Gas Reservoirs in Mrk\,590]{ALMA Probes the Molecular Gas Reservoirs in the \\ Changing-Look Seyfert Galaxy Mrk\,590}

\author[J. Y. Koay et al.]{
J. Y. Koay,$^{1}$\thanks{E-mail: koayjy@dark-cosmology.dk}
M. Vestergaard,$^{1,2}$
V. Casasola,$^{3,4}$
D. Lawther$^{1}$
and B. M. Peterson$^{5,6}$
\\
$^{1}$Dark Cosmology Centre, Niels Bohr Institute, University of Copenhagen, 2100 Copenhagen \O, Denmark\\
$^{2}$Steward Observatory, University of Arizona, Tucson, AZ 85721, USA\\
$^{3}$INAF - Osservatorio Astrofisico di Arcetri, 50125, Firenze, Italy\\
$^{4}$INAF - Istituto di Radioastronomia \& Italian ALMA Regional Centre, 40129 Bologna, Italy\\
$^{5}$Department of Astronomy, The Ohio State University, OH 43210, USA\\
$^{6}$Center for Cosmology and Astroparticle Physics, The Ohio State University, OH 43210, USA
}

\date{Accepted XXX. Received YYY; in original form ZZZ}

\pubyear{2015}

\begin{document}
\label{firstpage}
\pagerange{\pageref{firstpage}--\pageref{lastpage}}
\maketitle

\begin{abstract}

We investigate if the active galactic nucleus (AGN) of Mrk\,590, whose supermassive black hole was until recently highly accreting, is turning off due to a lack of central gas to fuel it. We analyse new sub-arcsecond resolution ALMA maps of the $^{12}$CO(3--2) line and 344\,GHz continuum emission in Mrk\,590. We detect no $^{12}$CO(3--2) emission in the inner 150\,pc, constraining the central molecular gas mass to $M({\rm H_2}) \lesssim 1.6 \times 10^5\, {M_{\odot}}$, no more than a typical giant molecular gas cloud, for a CO luminosity to gas mass conversion factor of $\alpha_{\rm CO}\sim 0.8\,{M_{\odot}\,\rm (K \,km\,s^{-1}\,pc^{2}})^{-1}$. However, there is still potentially enough gas to fuel the black hole for another $2.6 \times 10^5$ years assuming Eddington-limited accretion. We therefore cannot rule out that the AGN may just be experiencing a temporary feeding break, and may turn on again in the near future. We discover a ring-like structure at a radius of $\sim 1$\,kpc, where a gas clump exhibiting disturbed kinematics and located just $\sim 200$\,pc west of the AGN, may be refueling the centre. Mrk\,590 does not have significantly less gas than other nearby AGN host galaxies at kpc scales, confirming that gas reservoirs at these scales provide no direct indication of on-going AGN activity and accretion rates. Continuum emission detected in the central 150\,pc likely originates from warm AGN-heated dust, although contributions from synchrotron and free-free emission cannot be ruled out.

\end{abstract}


%%%%%%%%%%%%%%%%%%%%%%%%%%%%%%%%%%%%%%%%%%%%%%%%%%

%%%%%%%%%%%%%%%%% BODY OF PAPER %%%%%%%%%%%%%%%%%%

\section{Introduction}\label{introduction}

Active galactic nuclei (AGNs) are associated with the accretion of gas onto supermassive black holes residing in the centres of galaxies. The accretion-disk, feeding the black hole, generates continuum photons that photoionize and excite a gaseous region (known as the broad-line region) located just beyond the disk, light-days to light-weeks away from the black hole. This region produces the characteristic broad emission lines with velocity widths of $v >$\,2000\,km\,s$^{-1}$ (induced by the gravity of the black hole) observed in the optical-UV spectra of a class of AGNs (the so-called `Type 1 AGNs'). The photoionization of lower-velocity gas at distances of hundreds of parsecs by the accretion-disk photons produces narrower emission lines ($v <$\,1000\,km\,s$^{-1}$). To explain the absence of broad emission lines in the spectra of a fraction of AGNs (known as `Type\,2 AGNs'), AGN structural models \citep{antonucci93,urrypadovani95} posit that the accretion disk and broad-line region are surrounded by a dusty torus that obscures our view of these central regions in objects where the accretion disk is viewed (near) edge on. Only the narrow-line emission originating from more extended regions not obscured by the torus can be observed in these Type\,2 AGNs.

The occurrence of AGN activity in a galaxy is contingent upon the availability of gas to fuel the black hole. However, the mechanisms by which cold gas is transported from the galaxy disk at kpc scales to hundreds of pc scales, and further down into the inner few parsecs to sustain or trigger AGN activity, remain poorly understood to this day \citep[e.g.,][and references therein]{alexanderhickox12}. The interplay between gas inflows and outflows on these scales, as well as the presence (or absence) of dynamical barriers, determine the efficiencies at which the central black hole is fueled. These processes also influence the duty cycles of AGNs, i.e., the typical timescales during which AGNs are turned on (and off). Integral Field Unit (IFU) observations of nearby AGNs have revealed inflows of ionized gas down to 10\,pc scales \citep[e.g.][]{fathietal06,storchi-bergmannetal07, riffeletal08, schnorrmulleretal11}. However, the estimated inflow rates of the ionized gas in these observations are much lower than the observed black hole accretion rates. This is because the gas is predominantly molecular near the centre of galaxies \citep[e.g.,][]{walteretal08,bigieletal08}, such that ionized gas becomes a poor tracer of inflowing gas in the centre. The NUclei of GAlaxies (NUGA) program \citep{garcia-burilloetal03} has observed the distribution of cold molecular gas (as traced by the presence of carbon monoxide, CO) in 16 Seyfert and low-ionization nuclear emission-line region (LINER) galaxies at cosmic distances of 4\,Mpc to 43\,Mpc with the IRAM Plateau de Bure Interferometer (PdBI), achieving a spatial resolution of about 100\,pc. However, evidence for inflowing gas has been found only in one third of the NUGA galaxies on these scales \citep{lindt-kriegetal08, huntetal08, casasolaetal08, casasolaetal11, garcia-burilloetal09, garcia-burillocombes12}. The non-detection of direct evidence of fueling in the other two thirds of the NUGA galaxies may be due to outflowing gas driven by nuclear star formation, AGN winds and/or AGN jets that temporarily disrupt the nuclear fueling. In such a scenario, AGNs will undergo successive periods of fueling and starvation. More recent observations with the Atacama Large Millimetre/submilllimetre Array (ALMA) are probing the gas fueling mechanisms of AGNs at even smaller spatial scales, thereby mapping the molecular interstellar medium at 25\,pc to 100\,pc scales in local AGN host galaxies, including NGC\,1433 \citep{combesetal13}, NGC\,1566 \citep{combesetal14} and NGC\,1068 \citep{garcia-burilloetal14}. These studies reveal gas inflows and outflows occuring simultaneously on scales of tens of pc. At scales smaller than 10\,pc, the physical connections and coupling of energetics between the different gas components, including the accretion disk gas, the broad-line emitting gas, the narrow-line emitting gas and the cold molecular gas, are still unknown.  

Between the years 2006 and 2012, the broad H$\beta$ emission line in the optical spectrum of the AGN in Mrk\,590, a Seyfert galaxy, has disappeared. Examination of 40 years of published data on Mrk\,590 reveal an increase in the optical-UV continuum, broad-line and narrow-line fluxes from the 1970s to the early 1990s \citep[see][and references therein for a complete account]{denneyetal14}. Since then, the strengths of the broad emission lines (including H$\alpha$ and C\,\textsc{iv}) have steadily decreased to their present weak states. In the period between the early 1990s to the present, the X-ray continuum flux has also weakened, while the nuclear optical-UV continuum emission has faded to the point where the observed continuum can be fully accounted for by stellar population models of the host galaxy \citep{denneyetal14}. While still clearly visible, the narrow $[\rm O\, \textsc{iii}]\, \lambda 5007$ line has also decreased in flux by a factor of $\sim2$. The weakening of the $[\rm O\, \textsc{iii}]$ line appears to be delayed by up to ten years relative to that of the optical-UV continuum and broad H$\beta$ line, consistent with the larger spatial extent of the narrow-line region. The physical properties of Mrk\,590 and its AGN are listed in Table~\ref{Mrk590properties}.

\begin{table}
\centering
\caption{General properties of the Seyfert galaxy Mrk\,590 and its nucleus.}
\label{Mrk590properties}
\begin{tabular}{lcl}
\hline
\hline
Property & Value & Ref.  \\
\hline
R.A. (J2000)   		& 02$\rm ^h$ 14$\rm ^m$ 33.5$\rm ^s$	& 1 \\
Dec (J2000)			& $-$00$^{\circ}$ 46$^{\prime}$ 00$^{\prime\prime}$		& 1 \\
Morphology 			& 			SA(s)a 			& 1 \\
Galaxy inclination		   	&			$25^{\circ}$    & 2\\
Galaxy major axis	 P.A.			& 			$-55^{\circ}$ & 3\\
Redshift				&			0.0264			& 1 \\
Luminosity distance, ${D_{\rm L}}^a$ & 115.4\,Mpc & ...\\
Linear scale$^a$	        &  		1$^{\prime\prime} = 531\,$pc & ...\\
$v_{\rm sys}$ (barycentric)$^b$&	$\rm 7910\, km\,s^{-1}$ & 4\\
HI velocity			&	$\rm 7910\, km\,s^{-1}$ & 4$^d$\\
HI FWZI$^c$ & $\rm 380\, km\,s^{-1}$ & 4$^d$\\
Total HI mass				&	$30 \times 10^9 M_{\odot}$ & 4$^d$\\
$^{12}$CO(1--0) velocity & $\rm 7945\, km\,s^{-1}$ & 5$^d$\\
$^{12}$CO(1--0) FWHM & $\rm 205\, km\,s^{-1}$ & 5$^d$\\
$[\rm O\, \textsc{iii}]\, \lambda 5007$ velocity & 		$\rm 7950\, km\,s^{-1}$	& 6$^d$\\
$[\rm O\, \textsc{iii}]\, \lambda 5007$ FWHM & $\rm 400 \, km\,s^{-1}$ & 6$^d$\\
Black Hole Mass  & $4.75\pm 0.74 \times 10^7\,M_{\odot}$	&  7 \\
$L_{\rm bol}$ (1990s)$^e$ & $ \rm \sim 5.8 \times 10^{44}\,erg\,s^{-1}$ & 7\\
$L_{\rm bol}$ (2013)$^e$ & $\rm \sim 3.4 \times 10^{42}\,erg\,s^{-1}$ & 8\\

\hline
\end{tabular}
\begin{flushleft}
$^a$Values are derived from the source redshift using our adopted cosmology.\\
$^b$The systemic velocity of the galaxy relative to the Solar System barycentre.\\
$^c$FWZI is the full width of the emission line at zero intensity. \\
$^d$The uncertainties of the specified values are not quoted in the original papers.\\
$^e$The AGN bolometric luminosity, $L_{\rm bol}$, values are estimates and therefore no errors are not quoted in the original papers.\\
References. -- (1)~NASA Extragalactic Database, NED. (2)~\citet{whittle92}. (3)~\citet{schmittkinney00}. (4)~\citet{heckmanetal78}. (5)~\citet{maiolinoetal97}. (6)~\citet{vrtilekcarleton85}. (7)~\citet{petersonetal04}. (8)~\citet{denneyetal14}.
\end{flushleft}
\end{table}

One possible explanation for the fading of the AGN emission in Mrk\,590 is that the accretion disk and broad-line region are presently obscured by a, perhaps transiting, intervening cloud of optically thick gas. Variable obscuration can occur if the dusty torus, posited in AGN models, has a clumpy distribution \citep[e.g.,][]{elitzur12}. Variable X-ray absorption observed in the so-called `changing-look' AGNs \citep[e.g.,][]{bianchietal05} are thought to arise due to such clumpy obscuring clouds of gas and dust. However, \citet{denneyetal14} argue from the observed variability of the $[\rm O\, \textsc{iii}]\, \lambda 5007$ line that obscuration is unlikely to be responsible for the observed changes in Mrk\,590, since the obscuring cloud must cover both the continuum source along our line-of-sight, \textit{and} as viewed from the more extended narrow-line region as well. The more likely explanation for this behavior in Mrk\,590, is that the AGN is now accreting at a much lower rate, and may be close to turning off. This would result in a lack of ionizing continuum photons to sustain the broad-line emission, as modeled by \citet{elitzuretal14}. The dramatic changes observed in Mrk\,590 can therefore occur if the black hole has accreted all the cold gas in its vicinity, or if the gas has been ejected via outflows \citep[e.g.,][]{progakurosawa10}, and is not replenished by cold gas inflowing from the galaxy. In other words, the AGN in Mrk\,590 may be turning off due to very little gas being left in the centre of the galaxy. If no further gas is transported into the nucleus, the black hole is permanently starved and thereby turns quiescent, joining the majority of the nearby galaxy population. 

This abrupt weakening of the AGN in Mrk\,590 is surprising and unexpected, considering that it was accreting at $\sim$10\% of the Eddington limit in the 1990s \citep{petersonetal04}. The accretion rate has since decreased by two orders of magnitude to 0.05\% Eddington by 2013. While AGNs are indeed expected to turn off their activity as the black hole runs out of gas, this is believed to occur gradually over very long time-scales of several to hundreds of millions of years until the activity fades below our detection limit. This is based, in part, on observations of a large population of low-luminosity active galaxies with a wide range of activity levels \citep[e.g.,][]{baldwinetal81}. The disappearance of the H$\beta$ line also demonstrates that the absence of broad emission lines is not necessarily attributable only to source inclination, as posited by AGN unification schemes \citep{antonucci93,urrypadovani95}. While there is no doubt that orientation and the properties of the obscuring torus are the primary factors affecting the strength of the broad emission lines in AGN spectra, as evidenced by the detection of broad emission lines in polarized light in some Type\,2 AGNs \citep[e.g.,][]{antonuccimiller85}, this is clearly not the only explanation. The absence of polarized broad emission lines in a subset of Type\,2 AGNs have prompted suggestions that their lack of broad lines may not necessarily be due to an obscured broad line region but may be due to a lack of actual broad emission line gas \citep[e.g.,][]{tran01,tran03,panessabassani02,laor03}.

While the broad-emission lines of AGNs are known to be variable in flux \citep[][]{peterson88}, such dramatic changes in line strengths as seen in Mrk\,590 are very rare, and have been observed in only a handful of AGNs and LINERs. In fact, besides Mrk\,590, the complete disappearance of the broad H$\beta$ emission line has only been observed in four other AGNs to date. In one such object, NGC\,7603, the H$\beta$ line disappeared in a span of a year \citep{tohlineosterbrock76}. In another Seyfert galaxy, NGC\,4151, spectra observed in 1984 revealed the disappearance of the broad components of the H$\beta$ and H$\gamma$ Balmer lines that were detectable until 1983, although strong narrow components of these lines remained \citep{penstonperez84}. Further monitoring of the continuum and Balmer line fluxes of NGC\,4151 have shown continued variability since then, and the AGN returned to its previous active state by 1990 \citep{oknyanskiietal91}. The disappearance of the H$\beta$ line within a span of a decade was also recently reported for the $z = 0.31$ quasar SDSS\,J015957.64+003310.5 \citep{lamassaetal15} and the $z=0.246$ quasar SDSS J101152.98+544206.4 \citep{runnoeetal15}. The opposite behavior has also been observed. The sudden appearance or strengthening of the broad emission lines (within timescales of years and decades) has been observed in Mrk\,1018 \citep{cohenetal86}, NGC\,1097 \citep{storchi-bergmannetal93}, NGC 7582 \citep{aretxagaetal99}, and NGC\,2617 \citep{shappeeetal14}. These objects exhibiting large fluctuations in the broad emission line strengths have been referred to as `optical changing-look AGNs' \citep[e.g.,][]{lamassaetal15}.

Such rare and extreme changes in the strength of the optical-UV emission in AGNs provides excellent opportunities for studying the physics of black hole accretion and the coupling of energetics between the various AGN components emitting at radio to X-ray wavelengths. Additionally, it enables us to study the physics of AGN fueling and outflows.

To test the hypothesis that the nucleus of Mrk\,590 has run out of gas to fuel it, we obtained ALMA observations of the $^{12}$CO(3--2) emission line to trace the molecular gas in the circumnuclear regions of Mrk\,590. The combination of high angular resolution and high sensitivity observations provided by ALMA enable us to measure the mass of the molecular gas (or place very strong constraints on it) in the centre of Mrk\,590. Prior to this, the only observation of CO gas in Mrk\,590 is a single dish detection of the $^{12}$CO(1--0) line with the NRAO 12\,m telescope, with a beam size of 55$''$ \citep{maiolinoetal97}. At a redshift of 0.0264 (1$'' = 531$\,pc), Mrk\,590 is a few factors more distant than the nearby galaxies probed by NUGA. With the sub-arcsecond angular resolution of ALMA, we can probe the molecular gas distributions and (sub-)mm continuum on scales of $\sim$100\,pc, comparable to that of the NUGA survey of the more nearby AGNs. With the flexible tuning capabilities of ALMA's correlator, we are also able to simultaneously observe the HCO$^+$(4--3) emission line as a tracer of denser gas expected closer to the black hole. These ALMA data are complemented by Submillimetre Array (SMA) observations of the $^{12}$CO(2--1) emission line, albeit at a lower angular resolution of $\sim 3''$. The main goals of these observations are (1) to examine the distribution of molecular gas in Mrk\,590 to determine if it has indeed run out of gas in the central 100\,pc as hypothesized, (2) to determine if Mrk\,590 on larger kpc scales also has less gas compared to other nearby Seyfert galaxies, (3) to determine from the CO gas kinematics if gas is being transported from kpc scales into the central 100\,pc to refuel the centre, and (4) to detect signatures of molecular outflows (e.g., components with high velocities in excess of the rotational velocities of the galaxy) that may potentially be depleting the gas in the centre of Mrk\,590. 

We present here the first interferometric observations of Mrk\,590 at (sub-)mm wavelengths, providing the first ever look into the distribution and kinematics of molecular gas in the inner kpc of this changing-look AGN. In Section~\ref{obsdata}, we describe the ALMA and SMA observations and data processing. We present the continuum maps, $^{12}$CO(3--2) line maps and $^{12}$CO(3--2) spectra in Section~\ref{results}. We also place upper limits on the strengths of the HCO$^+$(4--3) and $^{12}$CO(2--1) emission, which we do not detect. From the $^{12}$CO(3--2) emission, we derive estimates of the molecular gas masses and star formation rates at different spatial scales in Mrk\,590. We discuss in Section~\ref{discussion} the implications of our results on the gas reservoirs and nuclear fueling in Mrk\,590, plus the origin of the continuum emission. Section~\ref{conclusion} summarizes the main findings of this study. We adopt the following cosmology: $\Omega_m = 0.3$, $\Omega_\Lambda = 0.70$, and $H_0 = 70\,{\rm km\,s^{-1}\,Mpc^{-1}}$.

\section{Observations, Data Processing and Imaging}\label{obsdata}

	\subsection{ALMA Observations}\label{ALMAobs}
	
Observations were carried out with ALMA during Cycle\,2 on 2014 July 27. Mrk\,590 was observed with 31 antennas for a total duration of 1.8 hours on source, over two execution blocks. The phase tracking centre was set to $\alpha_{\rm J2000} = 02^{\rm h}14^{\rm m}33.579120^{\rm s}$, $\delta_{\rm J2000} =-00^{\circ}46'00\farcs27840$. The source was observed in Band 7 with four separate spectral windows centred at sky frequencies of 336.7\,GHz, 338.0\,GHz, 346.8\,GHz, and 348.7\,GHz. The spectral windows for observing the $^{12}$CO(3--2) and HCO$^+$(4--3) lines, centred at sky frequencies of 338.0\,GHz and 346.8\,GHz respectively, were configured to Frequency Division Mode (FDM) with channel widths of 0.488\,MHz and a total bandwidth of 1.875\,GHz per spectral window. The remaining two spectral windows were configured to Time Division Mode (TDM), with a channel width of 15.625\,MHz and a total usable bandwidth of 1.875\,GHz per spectral window for continuum observations.

Calibration was performed as part of the quality assurance by the Italian node of the European ALMA Regional Centre. The quasar J0238+166 was used to set the absolute flux scales of the target source. The quasars J0217+0144 (located 2$\fdg$6 away from Mrk\,590) and J0224+0659 were used as the complex gain calibrator and bandpass calibrator, respectively. Antenna DV11 was set as the reference antenna. We re-examined the calibrated data and calibration tables to ensure that all spurious data were flagged, and that there were no bad calibration solutions.  %problems with the calibrations. Data from 2 antennas were flagged (DV48 and DV07)
	
Further data processing and imaging were carried out using the Common Astronomy Software Applications (CASA) package \citep[version 4.2.1;][]{mcmullinetal07}. Examining the visibility amplitudes as a function of channel numbers, averaged over all baselines, reveals no detection of the HCO$^+$(4--3) line. `Dirty images' obtained at various channel widths (i.e., 10$\,\rm km\,s^{-1}$, 20$\,\rm km\,s^{-1}$, and 50$\,\rm km\,s^{-1}$) also show no evidence of HCO$^+$(4--3) line emission (as also evidenced in Figure~\ref{spectotal}). We therefore combined the data from the HCO$^+$ spectral window with that of the other two continuum spectral windows to produce the continuum images. We produced three separate images, for each image using a different weighting scheme: natural, Briggs (robustness $= 0.5$) or uniform weighting. Since the continuum images were produced from spectral windows centred at frequencies of 336.7\,GHz, 346.8\,GHz, and 348.7\,GHz, we define the mean frequency of 344\,GHz as the representative frequency for the continuum emission. Imaging and deconvolution were performed using the \texttt{clean} task in CASA. Components in the continuum images were cleaned iteratively down to a threshold of three times the image rms intensities, using a loop gain of 0.1. We generated images of 640 pixels $\times$ 640 pixels with pixel sizes of $0\farcs 05 \times 0\farcs 05$. The rms noise, beam sizes and beam orientations for all three continuum images are presented in Table~\ref{contprop}. 

\begin{table*}
\centering
\caption{Properties of the ALMA and SMA continuum images \label{contprop}}
\begin{tabular}{lccccccc}
\hline
\hline

\multicolumn{8}{l}{\textbf{ALMA (344\,GHz) -- Phase Centre: $\alpha_{\rm J2000} = 02^{\rm h}14^{\rm m}33.579120^{\rm s}$, $\delta_{\rm J2000} =-00^{\circ}46'00\farcs27840$}} \\
\hline
Weighting & rms & Beam size$^a$ & Beam P.A.  & Source size$^b$  & Source P.A.  & $S_{\rm int}$ & $S_{\rm peak}$ \\ 
& (mJy beam$^{-1}$) & ($'' \times ''$) & ($^{\circ}$) & ($'' \times ''$) & ($^{\circ}$) & (mJy) & (mJy beam$^{-1}$) \\
\hline
Natural &  0.031 & 0.35 $\times$ 0.29 & $-$69.42 & $<$ 0.25 $\times$ 0.09 &  ... & 0.733 $\pm$ 0.037 & 0.622 $\pm$ 0.031\\
Briggs & 0.036 & $0.34 \times 0.24$ & $-$67.45 & $0.263 \times 0.069$ & $93.0^{\circ}$ & $0.830 \pm 0.048$ &$0.621 \pm 0.036$\\
Uniform  & 0.109 & $0.32 \times 0.19$ & $-$70.30 & $<$ 0.22 $\times$ 0.10 & ... & 0.596 $\pm$ 0.101 & 0.656 $\pm$ 0.111 \\ 
\hline
\hline
\multicolumn{8}{l}{\textbf{SMA (219\,GHz) -- Phase Centre: $\alpha_{\rm J2000} = 02^{\rm h}14^{\rm m}33.561573^{\rm s}$, $\delta_{\rm J2000} =-00^{\circ}46'00\farcs09007$}} \\
\hline
Weighting & rms & Beam size$^a$ & Beam P.A.  & Source size$^b$  & Source P.A.  & $S_{\rm int}$ & $S_{\rm peak}$ \\ 
& (mJy beam$^{-1}$) & ($'' \times ''$) & ($^{\circ}$) & ($'' \times ''$) & ($^{\circ}$) & (mJy) & (mJy beam$^{-1}$) \\
\hline
Natural &  0.41 & 3.68 $\times$ 3.22 & 72.22 &  ... &  ... & 3.23 $\pm$ 0.31 & 4.38 $\pm$ 0.41\\
Briggs &  0.56 & 3.46 $\times$ 2.68 & 86.62 &  ... &  ... & 4.08 $\pm$ 0.44 & 5.22 $\pm$ 0.56\\
Uniform &  1.3 & 3.45 $\times$ 2.04 & $-$88.03 & ... &  ... & 6.9 $\pm$ 1.6 & 5.7 $\pm$ 1.3\\
\hline
\end{tabular}
\begin{flushleft}
$^a$The beam sizes are given as the full width at half maximum (FWHM) of the synthesized (clean) beam.\\
$^b$The estimated source sizes are beam-deconvolved.
\end{flushleft}
\end{table*}   

For the $^{12}$CO(3--2) line images, channels without line emission within the spectral window were used to model the continuum emission. The continuum emission model was then subtracted from the visibilities using the task \texttt{uvcontsub} in CASA. Imaging was done in $20\,{\rm km\,s}^{-1}$ channels using natural weighting. We applied uv-tapering at a lengthscale (uv-distance) of $300,000$ times the central observed wavelength of the CO spectral window, for better sensitivity to larger scale structures and to improve surface brightness sensitivity. As with the continuum images, we generated an image of 640 pixels $\times$ 640 pixels, with pixel sizes of $0\farcs 05 \times 0\farcs 05$. We obtain a synthesized beam size of 0$\farcs$47 $\times$ 0$\farcs$42 at a position angle, P.A., of $-76^{\circ}$. For the deconvolution, we applied a mask generated by selecting pixels with intensities greater than $4 \sigma$ (with $\sigma$ calculated over the entire image) in each channel. We then cleaned components within the mask for 4000 iterations with a gain of 0.05. Pixels with intensities greater than $4 \sigma$ in the residual channel images were then selected as new regions which were then added to the existing mask to form a new mask. Thereafter, we cleaned the components within the new mask for another 4000 iterations. This process was repeated five times, after which there were no longer any pixels with intensities greater than $4 \sigma$ in the residual channel images. Thereafter, we applied primary beam corrections to the image intensities using the task \texttt{impbcor}. The continuum and $^{12}$CO(3--2) line maps are presented and discussed further in Sections~\ref{resultscont} and \ref{resultsCO} respectively.

\subsection{SMA Observations}\label{SMAobs}
	
The SMA observations were carried out using the 230\,GHz receiver over two tracks of $\sim$12\,hours each, the first on 2014 October 1, and the second on 2014 October 2. The phase centre of the observations was set to $\alpha_{\rm J2000} = 02^{\rm h}14^{\rm m}33.561573^{\rm s}$, $\delta_{\rm J2000} =-00^{\circ}46'00\farcs09007$. The 18th chunk of the upper side-band was centred at a sky frequency of 224.6 GHz to observe the $^{12}$CO(2--1) line. This allowed the $^{13}$CO(2--1) and $^{12}$C$^{18}$O(2--1) lines to fall within the lower side-band. The spectral setup was configured such that there are 32 channels per chunk, with a channel width of 3.25\,MHz. The total bandwidth is 4\,GHz for each side-band. The target was observed for a total duration of 18.47 hours on-source with only seven out of eight antennas. Due to technical issues, antenna SMA3 was down for the entire duration of both tracks. %We note that observations during the first track were rated as `unsatisfactory' by SMA observers, due to low atmospheric transmission, but the data are still useable.
	
The SMA data were converted to FITS-IDI files using the Python script \texttt{sma2casa.py} and then to `.ms' format using the \texttt{smaImportFix.py} script\footnote{Both Python scripts were developed by Ken Young; available at: https://github.com/kenyoung/sma2casa} to enable us to perform the data processing using the CASA software package. We flagged the highest and lowest 10\% of the 32 channels in each chunk during this conversion to remove channels with very low gains at the chunk edges. We inspected the visibilities to look for bad data (e.g., outliers and phase jumps), which were subsequently flagged and removed. 

We calibrated the data from each track independently, using standard methods, with quasar 3C454.3 as the bandpass calibrator and the quasar J0224+0659 (located 8$^\circ$ away from Mrk\,590) as the complex gain calibrator. We used the planet Uranus to set the absolute flux scales. Antenna SMA4 was selected as the reference antenna. We then concatenated the data from the two tracks using the task \texttt{concat} in CASA. 

We found no significant detection of the $^{12}$CO(2--1), $^{13}$CO(2--1) and $^{12}$C$^{18}$O(2--1) lines in the SMA spectra. We generated three continuum images using natural, Briggs (robustness = 0.5) and uniform weighting respectively, and cleaned iteratively (with a loop gain of 0.1) down to a threshold of three times the image rms intensities. Visibilities from both the upper and lower side-bands were concatenated to achieve a total bandwidth of 8\,GHz for the maximum possible continuum sensitivity. The representative continuum frequency is 219\,GHz. Each image contains $\rm 256\, pixels \times 256\,pixels$ with a pixel size of $0\farcs 5 \times 0\farcs 5$. Image properties are provided in Table~\ref{contprop} and presented in Section~\ref{resultscont}.

\subsection{Ancillary Images}\label{ancdata}

We describe here additional archival radio and optical images of Mrk\,590 which we use to compare the distribution of the (sub-)mm continuum and $^{12}$CO(3--2) emissions relative to other physical components emitting at radio and optical wavelengths. 

\subsubsection{Very Large Array Continuum Image}\label{NVASimage}

We extracted 8.4\,GHz Very Large Array (VLA) continuum data of Mrk\,590 from the NRAO VLA Archive Survey (NVAS)\footnote{The NVAS can [currently] be browsed at the following website: http://archive.nrao.edu/nvas/}. The data are from VLA A-configuration observations carried out on 1991 June 25. We converted the AIPS pipeline-calibrated visibilities from its original UVFITS format into a `.ms' file using the CASA task \texttt{importuvfits}, to carry out the imaging with the CASA software. We generated an image of 640 pixels $\times$ 640 pixels, with a pixel size of $0\farcs05 \times 0\farcs05$. Uniform weighting was used, such that the synthesized beam of $0\farcs26 \times 0\farcs24$ (P.A. $= 26.3^{\circ}$) is comparable with that of our ALMA images. We cleaned components in the image iteratively (with a loop gain of 0.1) down to a threshold of three times the image rms intensity, using the \texttt{clean} task in CASA. The VLA image is discussed in Section~\ref{resultscont} along with our ALMA continuum image.

\subsubsection{Hubble Space Telecope Image}\label{HSTimage}

We downloaded the \textit{Hubble Space Telescope (HST)} Advanced Camera for Surveys (ACS) image of Mrk\,590 from the Hubble Legacy Archive\footnote{http://hla.stsci.edu}, observed on 2003 December 18 with an exposure time of 1020\,s using the F550M (\textit{V}-band) filter (proposal ID: 9851). L. Slavcheva-Mihova kindly provided a structure map produced from this image as presented by \citet{slavcheva-mihovamihov11}. We applied a coordinate shift of $00^{\rm h}00^{\rm m}00.044^{\rm s}$ and $-00^{\circ}00'00\farcs02$ to the \textit{HST}/ACS image and the structure map, so that the brightest ACS pixel in the nuclear region (at $\alpha_{\rm J2000} = 02^{\rm h}14^{\rm m}33.604^{\rm s}$, $\delta_{\rm J2000} =-00^{\circ}46'00\farcs17$, where the AGN is likely located) coincides with the new location of the galaxy centre derived from the ALMA 344\,GHz continuum image (described below in Section~\ref{resultscont}). We present both the \textit{HST}/ACS F550M image and the structure map in Section~\ref{resultsCO} together with the $^{12}$CO(3--2) line maps.

\section{Results}\label{results} 

\subsection{Continuum Maps}\label{resultscont} 

\begin{figure}
\begin{center}
\includegraphics[width=\columnwidth]{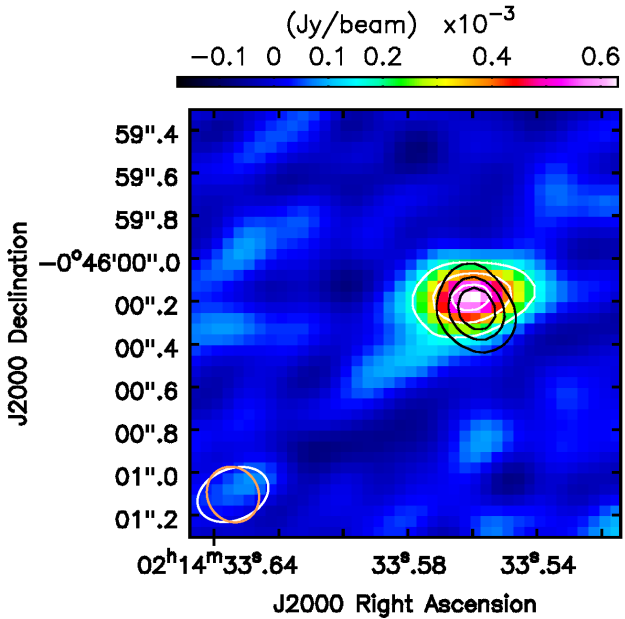}
\end{center}
\caption{{The ALMA continuum image at a mean central frequency of 344\,GHz, obtained using Briggs weighting (robustness = 0.5) and centred at the phase centre of the observations (Table~\ref{contprop}). The white contours show the 5$\sigma$, 10$\sigma$ and 15$\sigma$ (where 1$\sigma\, \sim$ 0.036 mJy\,beam$^{-1}$) continuum intensity levels. The FWHM ALMA synthesized beam ($0\farcs34 \times 0\farcs24$) is shown as the white ellipse at the bottom left corner of the image.  Black contours show the 8.4 GHz NVAS VLA image (described in Section~\ref{NVASimage}), with contour levels at 10$\sigma$, 20$\sigma$ and 30$\sigma$ (where 1$\sigma\, \sim$ 0.074 mJy beam$^{-1}$). The FWHM VLA synthesized beam ($0\farcs26 \times 0\farcs24$) is shown as the orange ellipse in the bottom left corner.} \label{contimage}}
\end{figure}

In all three images derived using the different weighting schemes, the ALMA continuum emission at 344\,GHz is detected as a single compact component close to the phase centre of the observations. The continuum map obtained using Briggs weighting is shown in Figure~\ref{contimage}, in which the source is detected at a $17\sigma$ significance. We used the CASA task \texttt{imfit} to fit a 2-dimensional Gaussian distribution to this compact component in all three images. We then extracted the source sizes (beam deconvolved), parallactic angles, integrated fluxes and peak fluxes (shown in Table~\ref{contprop}). The continuum emission is barely resolved when Briggs weighting is used (with slightly elongated emission in the east-west direction) and appears unresolved when natural and uniform weighting are used. The 344\,GHz continuum emission also spatially coincides with the unresolved 8.4\,GHz radio source in the NVAS image, as shown in Figure~\ref{contimage}, and thus is where the AGN most likely is located. Based on the Gaussian function fit to the Briggs-weighted continuum image, we define its centroid, at $\alpha_{\rm J2000} = 02^{\rm h}14^{\rm m}33.5605^{\rm s}$, $\delta_{\rm J2000} =-00^{\circ}46'00\farcs1851$, as the centre of the galaxy, as well as the location of the AGN. We discuss the origin of this central continuum emission in Section~\ref{contorigin}.

The $2\sigma$ jet-like feature protruding towards the south-east is most likely an image artifact. It appears only in the spectral window used for observing the HCO$^+$ line, which is slightly noisier than the other continuum spectral windows. Generating an image using an elongated, uv-tapered beam with a major axis parallel to the P.A. of this jet-like feature does not increase its signal-to-noise ratio as one would expect if the `jet' is real. Radio images at cm wavelengths also show no evidence of an extended jet, so this feature is unlikely to be astrophysical in origin.

\begin{figure}
\begin{center}
\includegraphics[width=\columnwidth]{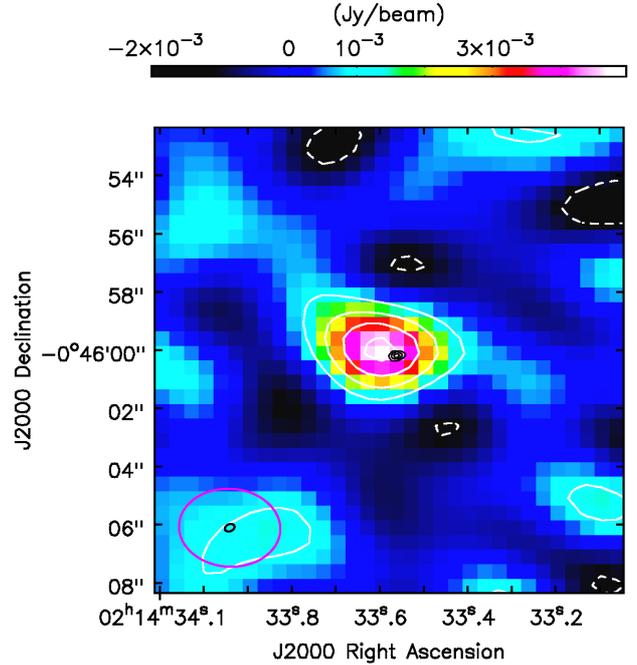}
\end{center}
\caption{{The SMA continuum image obtained using Briggs weighting (robustness = 0.5) at a central frequency of 219\,GHz, and centred at the phase centre of the observations (Table~\ref{contprop}). We obtain a 1$\sigma$ rms intensity of 0.56 mJy\,beam$^{-1}$. White contours represent $-2\sigma$, 2$\sigma$, 4$\sigma$, 6$\sigma$, and 8$\sigma$ levels, with negative levels shown dashed. The SMA FWHM synthesized beam of $3\farcs46 \times 2\farcs68$ is shown as the magenta ellipse at the bottom left corner of the image. Black contours show the ALMA 344\,GHz continuum image (Figure~\ref{contimage}), with contour levels at 5$\sigma$, 10$\sigma$ and 15$\sigma$ (where 1$\sigma\, \sim$ 0.036 mJy\,beam$^{-1}$). The FWHM ALMA synthesized beam of $0\farcs34 \times 0\farcs24$ is shown as the small black ellipse in the middle of the SMA beam ellipse in the bottom left corner.} \label{SMAcontimage}}
\end{figure}

The SMA continuum emission at 219\,GHz is also detected (at 9$\sigma$ significance in the Briggs-weighted image) as a single, marginally resolved component at the phase centre (Figure~\ref{SMAcontimage}). We note that the ALMA continuum emission is spatially offset by $\sim 0\farcs5$ from the pixel containing the peak flux in the SMA image. Nevertheless, their spatial locations are consistent to within the absolute astrometric uncertainties, which for each of these instruments can be as large as their corresponding synthesized beamwidths. The peak fluxes and integrated flux densities obtained by fitting a 2-dimensional Gaussian distribution (using \texttt{imfit}) to the continuum emission are shown in Table~\ref{contprop}. However, \texttt{imfit} is unable to deconvolve the clean beam for source size estimates, due to the source being marginally resolved in only one direction.

\subsection{$^{12}$CO(3--2) Maps and Spectra}\label{resultsCO} 

\begin{figure*}
\begin{center}
\includegraphics[width=\textwidth]{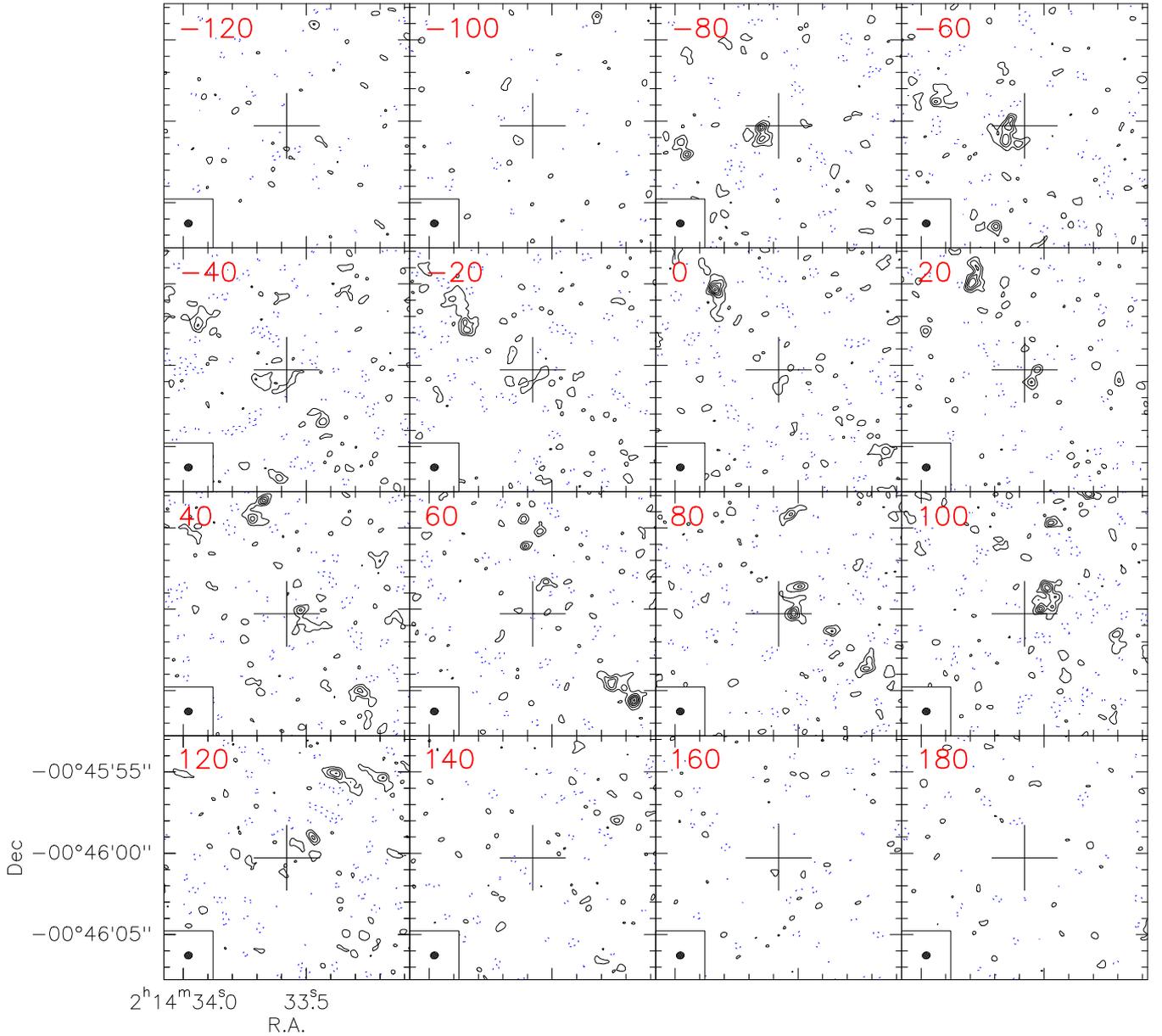}
\end{center}
\caption{{ALMA $^{12}$CO(3--2) channel maps of the inner 15$''$ by 15$''$ of Mrk\,590, centred on the phase centre of the observations (Table~\ref{contprop}), generated using natural weighting and uv-tapering at a length-scale 300,000 times the central observed wavelength. Each channel has a width of $\rm 20\,\,km\,s^{-1}$. The central velocity of the channel, relative to the systemic HI emission velocity ($\rm 7910\,km\,s^{-1}$, \,Table~\ref{Mrk590properties}) is labeled in units of km\,s$^{-1}$ in each panel. The contours correspond to $-2\sigma$, $2\sigma$, $4\sigma$, $6\sigma$, $8\sigma$, $10\sigma$ and $12\sigma$ intensity levels, where 1$\sigma \sim 0.5\, \rm mJy\,beam^{-1}$ for each channel in the residual maps. Negative intensity levels are shown as dotted (blue) contours. We obtain a spatial resolution of $0\farcs 47 \times 0\farcs 42$ (FWHM beam shape shown in bottom left corner). No primary beam corrections are applied to the image intensities. } \label{channelmaps}}
\end{figure*}

\begin{figure*}
\begin{center}
\includegraphics[width=\textwidth]{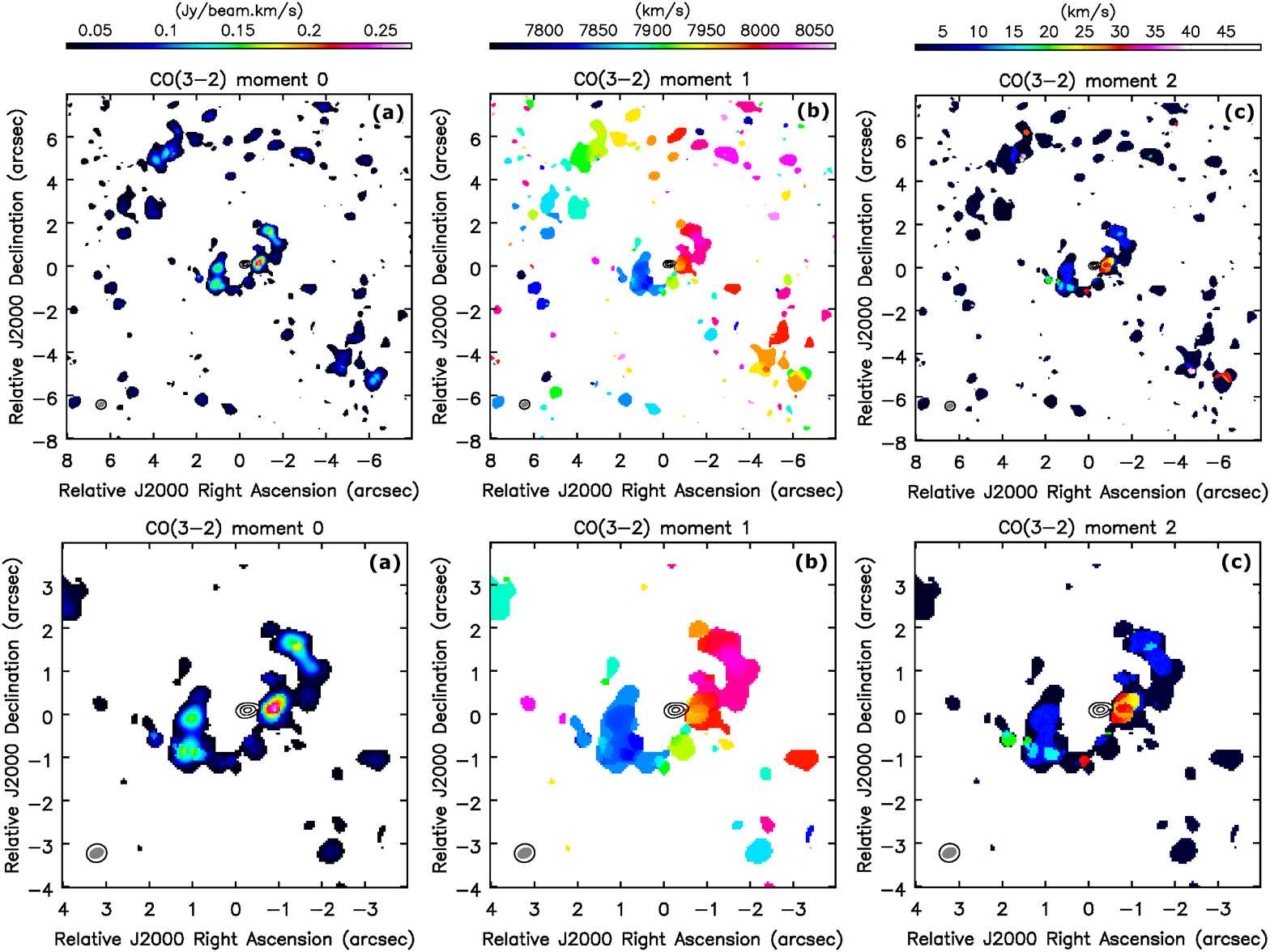} 
\end{center}
\caption{{ALMA maps of the (a) integrated fluxes (moment 0), (b) intensity weighted velocity fields (moment 1) and (c) velocity dispersions (moment 2) of the $^{12}$CO(3--2) line in Mrk\,590, derived from the channel maps in Figure~\ref{channelmaps} (FWHM beam size: 0\farcs47 $\times$ 0\farcs42, with no primary beam corrections). The panels in the top row show the inner 16$''$ by 16$''$ ($\rm \sim 8.5\,kpc \times 8.5\,kpc$), while the bottom row shows the zoomed-in versions of the same maps, for the inner 8$''$ by 8$''$  ($\rm \sim 4.2\,kpc \times 4.2\,kpc$). Flux clipping was performed to set pixels with values less than 3$\sigma$ to zero, where 1$\sigma \sim 0.5\, \rm mJy\,beam^{-1}$ in each 20 $\rm km\,s^{-1}$ channel in the residual maps. Black contours show the ALMA 344\,GHz continuum image (Briggs weighting, robustness: 0.5; contour levels: 5$\sigma$, 10$\sigma$, 15$\sigma$; 1$\sigma \sim 0.036\,{\rm mJy\,beam^{-1}}$). Ellipses in the bottom left corner are the beam shapes for the continuum image (grey) and the $^{12}$CO(3-2) maps (black outline). Origo is the phase centre of the observations (Table~\ref{contprop}).} \label{momentmaps}}
\end{figure*}

In this section, we present the ALMA $^{12}$CO(3--2) maps and spectra, plus discuss the molecular gas distribution and morphology in the inner 10\,kpc of Mrk\,590. Figure~\ref{channelmaps} shows 16 of the $^{12}$CO(3--2) channel maps of 20$\rm\, km\,s^{-1}$ width, covering the range of velocities where the line emission is detected. These channel maps were generated using the GILDAS\footnote{http://www.iram.fr/IRAMFR/GILDAS} software package \citep{guilloteaulucas00}. No primary beam corrections are applied to these images. The strongest $^{12}$CO(3--2) emitting components in the channel maps are detected at $\sim 12\sigma$ significance. We use the task \texttt{immoments} in CASA to produce the moment maps (Figure~\ref{momentmaps}), corresponding to the integrated fluxes (moment 0), intensity weighted velocity fields (moment 1) and velocity dispersions (moment 2) of the $^{12}$CO(3--2) emission line. In generating these moment maps, we set pixels with less than 3$\sigma$ intensities in the channel maps to zero. Here and for the rest of the paper, the radial velocities ($v$) are defined using the optical convention such that $v = cz$ (relative to the rest frame of the Solar system barycentre), where $c$ is the speed of light and $z$ is the redshift of the emission line in question.  

We observe two ring-like CO gas structures (Figure~\ref{momentmaps}): an outer `circular ring' at a radius of $\sim 6''$ (3\,kpc) from the centre of the galaxy, and an inner `elliptical ring' at a radius of $\sim$2$^{\prime\prime}$ (1\,kpc) from the centre. However, there is no detection of $^{12}$CO(3--2) emission at the position where the 344\,GHz continuum emission is observed in the Briggs-weighted continuum image (with a synthesized beam size of 180\,pc $\times$ 120\,pc), the likely location of the AGN. There is therefore no significant $^{12}$CO(3--2) emission within the central $\sim 150$\,pc (taken as the mean of the major axis and minor axis of the synthesized beam) of Mrk\,590.   

\begin{figure*}
\begin{center}
\includegraphics[width=\textwidth]{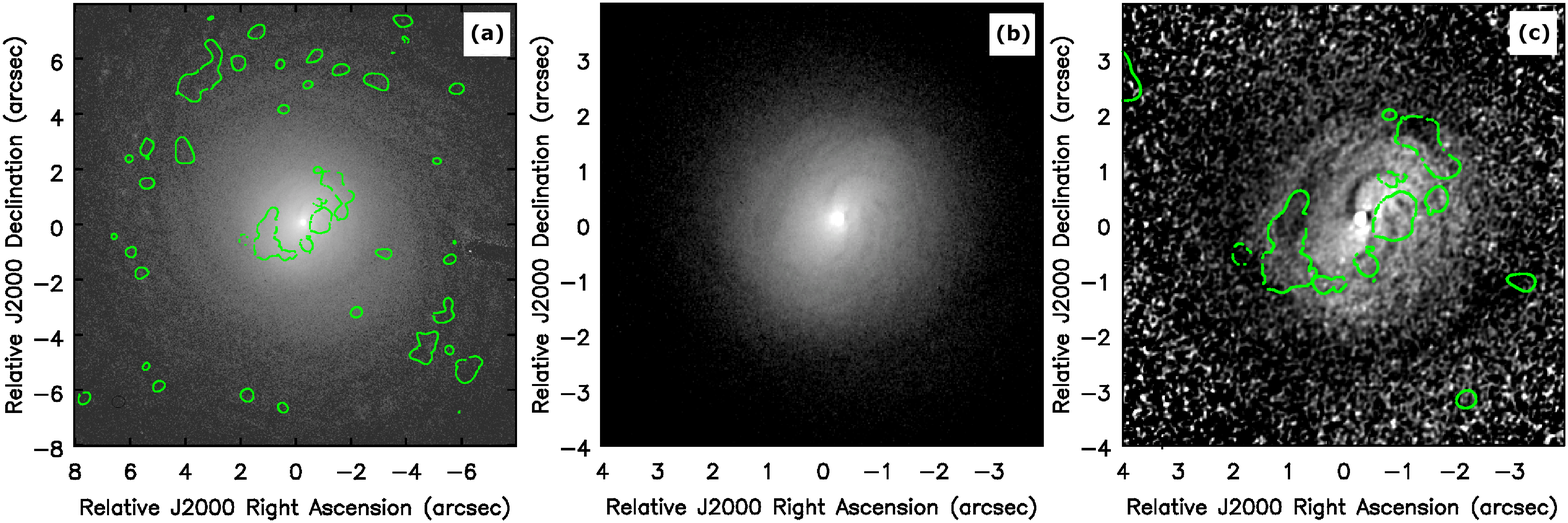}
\end{center}
\caption{{(a) \textit{HST}/ACS F550M (\textit{V}-band) image of the central 16$''$ by 16$''$ ($\rm \sim 8.5\,kpc \times 8.5\,kpc$) of Mrk\,590, with the $^{12}$CO(3--2) integrated flux contour map overlaid. The green contours denote flux levels equivalent to 15\% of the peak integrated flux (0.27\,Jy\,beam$^{-1}$\,km\,s$^{-1}$), and thus roughly outline the location of the detected $^{12}$CO(3--2) emission. (b) The zoomed-in \textit{HST}/ACS F550M image shows the dust lanes in the inner 8$''$ by 8$''$ of Mrk\,590. (c)~The $^{12}$CO(3--2) integrated flux contour map overlaid on the \textit{HST}/ACS F550M structure map presented by \citet{slavcheva-mihovamihov11} for the inner 8$''$ by 8$''$ ($\rm \sim 4.2\,kpc \times 4.2\,kpc$) of Mrk\,590. The outer CO ring corresponds to dusty spiral structures in the \textit{HST} image, while the inner ring appears to trace faint dust lanes in the central 4$''$; this is clearest in panel c. The clump west of the AGN (component C, Figure~\ref{spectregions}), overlaps with three faint dust lanes. Origo is the phase centre of the ALMA observations (Table~\ref{contprop}).} \label{hstoverlay}}
\end{figure*}

The outer ring-like structure coincides with spiral structures seen in the \textit{HST}/ACS F550M image shown in Figure~\ref{hstoverlay}a. The gas kinematics may be dominated by rotation, as indicated by the velocity fields in Figure~\ref{momentmaps}b. Since these structures are located close to the edges of the ALMA primary beam with slightly lower sensitivity, it is likely that only the brightest components are detected and revealed in the moment maps. Larger-scale extended structures would also be resolved out. We estimate the maximum recoverable scale to be $\sim 4''$ (2\,kpc) based on the shortest projected baselines (uv-distance $\sim$30,000 times the observing wavelength) of our ALMA observations.

\begin{figure*}
\begin{center}
\includegraphics[width=\textwidth]{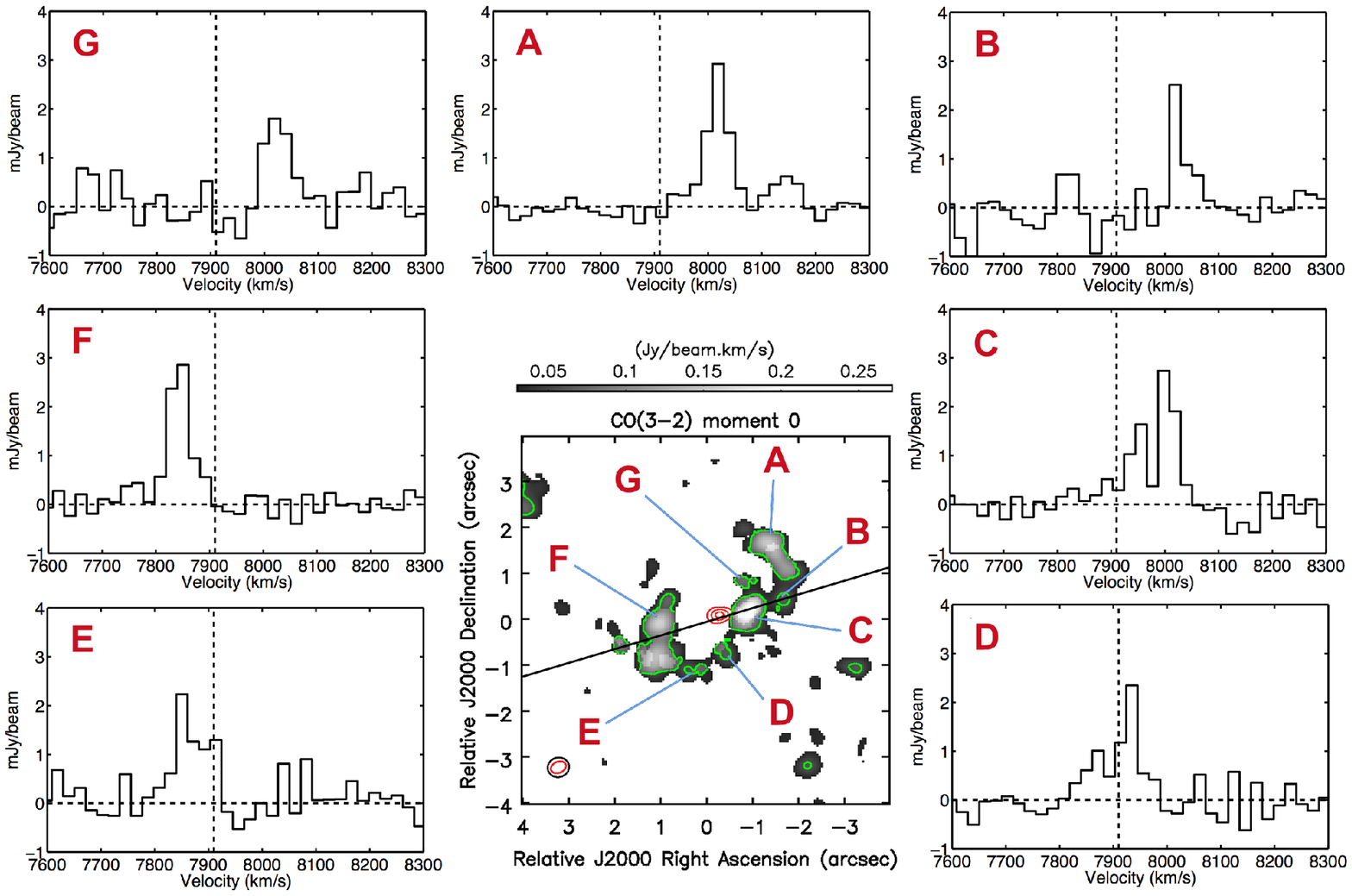}
\end{center}
\caption{{Mean $^{12}$CO(3--2) spectra, with primary beam corrected fluxes, of different components of the inner gas ring. The dashed vertical line shows the systemic HI emission line velocity (7910$\rm \,km\,s^{-1}$, Table~\ref{Mrk590properties}). The spectra of components A to G are extracted from the regions outlined by the green contours in the moment 0 map (central panel) of the inner 8$''$ by 8$''$ ($\rm \sim 4.2\,kpc \times 4.2\,kpc$) of Mrk\,590. The red contours in this map show the continuum flux, with contour lines at 5$\sigma$, 10$\sigma$ and 15$\sigma$ (1$\sigma \sim 0.036\rm \,mJy\,beam^{-1}$). The straight black line shows the position of the slit through which the position-velocity diagram in Figure~\ref{pv} is extracted.} \label{spectregions}}
\end{figure*}

The spectra (with primary beam corrections) of the different components of the inner gas `ring' are shown in Figure~\ref{spectregions}. The CO emission in this inner gas structure appears to trace faint dust lanes seen in the \textit{HST} \textit{V}-band image (Figure~\ref{hstoverlay}, panels b and c), possibly associated with circumnuclear star-forming regions in the central 2 kpc. \citet{poggemartini02} claim the presence of a nuclear stellar bar in Mrk\,590, based on an earlier lower resolution \textit{HST}/WFPC2 image. Structure maps from higher resolution \textit{HST}/ACS images \citep{slavcheva-mihovamihov11} do not show strong signatures of a stellar bar, but reveal what appear to be straight dust lanes similar to that observed at leading edges of stellar bars (Figure~\ref{hstoverlay}c). If indeed there is a nuclear stellar bar, its position angle would be roughly aligned with the major axis of this inner CO ring. Components A and F of the CO gas ring (Figure~\ref{spectregions}), both of which appear to trace the outer dust lanes, would be lying just outside this stellar bar. As with the outer ring, the channel maps (Figure~\ref{channelmaps}) and velocity fields (Figure~\ref{momentmaps}) show that gas velocities of this inner ring are consistent with rotation-dominated kinematics. Our crude, preliminary kinematics model (Section~\ref{kinematicmodels}) is also consistent with rotation-dominated gas in the inner ring, with signatures of non-rotating velocity components in region C (Figure~\ref{spectregions}). 

\begin{figure}
\begin{center}
\includegraphics[width=\columnwidth]{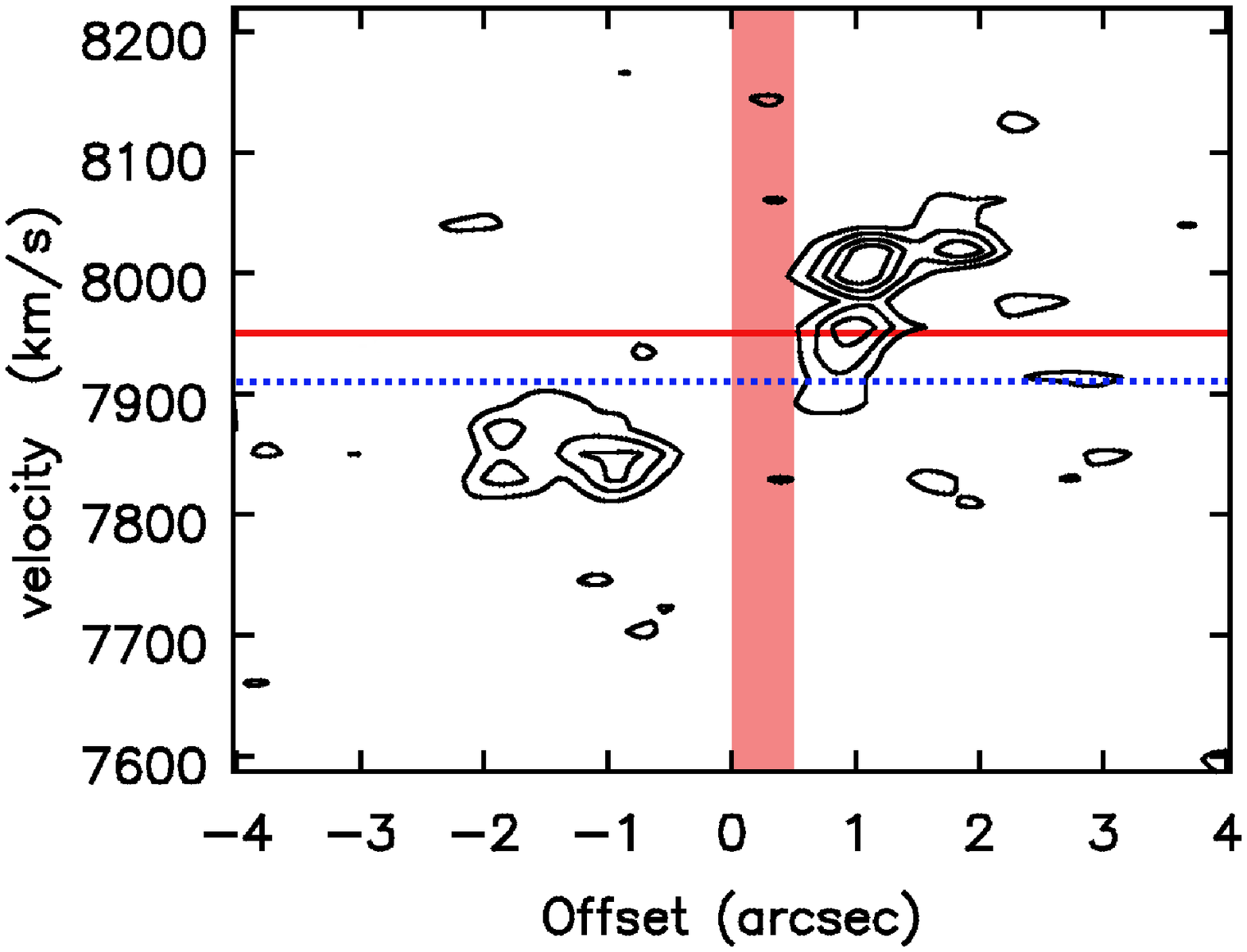}
\end{center}
\caption{{Position-velocity (pv) diagram of the $^{12}$CO(3--2) line, obtained through a 1-pixel wide slit at P.A. = 107$^{\circ}$ (grey line in Fig.~\ref{spectregions}). The X-axis shows the offset relative to the ALMA phase centre. Contour levels are 20\%, 40\%, 60\% and 80\% of the peak flux (4.0 mJy\,beam$^{-1}$) of the pv diagram. The red vertical bar denotes the location of the continuum emission. The horizontal (blue) dotted line shows the systemic HI emission velocity of 7910$\rm \,km\,s^{-1}$ for the galaxy (Table~\ref{Mrk590properties}). The horizontal (red) solid line indicates the centre of the $[\rm O\,\textsc{iii}]\,\lambda 5007$ emission line \citep{vrtilekcarleton85}} at 7950$\rm \,km\,s^{-1}$, likely the velocity of the AGN. The systemic velocity of 7945$\rm \,km\,s^{-1}$ as derived from the $^{12}$CO(2--1) line \citep{maiolinoetal97} is close to the centre of the $[\rm O\,\textsc{iii}]\,\lambda 5007$ line and is not shown.\label{pv}}
\end{figure}

The intriguing bright clump (component C, Figure~\ref{spectregions}) located $\sim$200\,pc west of the AGN coincides with three faint dust lanes in the \textit{HST} image (Figure~\ref{hstoverlay}, panels b and c). The $^{12}$CO(3--2) spectrum contains two strong velocity components (with possibly a third weaker component; Figure~\ref{spectregions}). It is unclear if they are rotating components that are simply unresolved, or comprise non-rotational components. The different components remain spatially unresolved in the images derived using Briggs weighting (robustness = 0.5) and uniform weighting. This bright clump is closest to the AGN spatially, with one of the velocity components (peaking at $\sim 7950\, \rm km\,s^{-1}$) comparable to the systemic velocity of Mrk\,590 as determined from the HI 21\,cm line (Table~\ref{Mrk590properties} and Figure~\ref{pv}). The velocity of the AGN $\left[ \rm O\, \textsc{iii} \right]\,\lambda 5007$ emission line is also similar to this gas component, indicating that this CO component is very likely close to the AGN in velocity-space. We further examine the kinematics of component C and the inner gas ring in Section~\ref{kinematicmodels}.

\begin{figure}
\begin{center}
\includegraphics[width=\columnwidth]{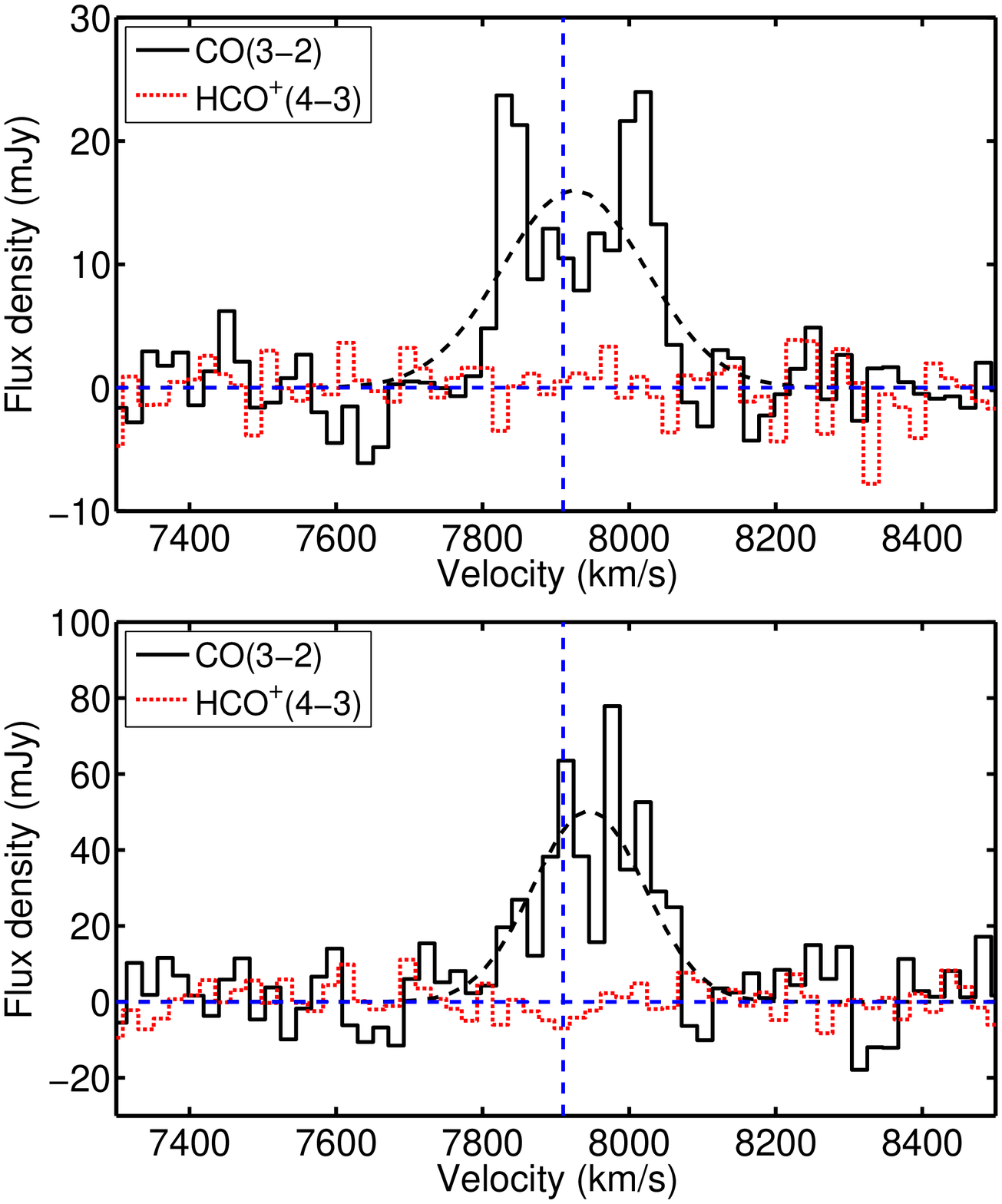}
\end{center}
\caption{{Spectra of $^{12}$CO(3--2) (solid black histogram) and HCO$^+$(4--3) (dotted (red) histogram, no detection) emission spatially integrated over the inner 4$^{\prime\prime}$ (top panel) and the full 18$^{\prime\prime}$ ALMA primary beam (bottom panel). The $^{12}$CO(3--2) profile is characteristic of a rotating disk. Channel widths are 20\,km\,s$^{-1}$. The black dashed curves show a single Gaussian function fitted to each of the $^{12}$CO(3--2) spectra. The dashed vertical lines indicate the systemic HI emission-line velocity (7910$\rm \,km\,s^{-1}$, Table~\ref{Mrk590properties})} \label{spectotal}}
\end{figure}

The $^{12}$CO(3--2) spectra integrated over the central 4$^{\prime\prime}$ (encompassing the inner gas ring) and over the entire 18$^{\prime\prime}$ ALMA primary beam are shown in Figure~\ref{spectotal}. Primary beam corrections are applied to the fluxes. By fitting single Gaussian functions to each spectrum, we obtain FWHM widths of $231 \pm 80 \, \rm km\,s^{-1}$ for the inner 4$^{\prime\prime}$ integrated spectrum and $190 \pm 60 \, \rm km\,s^{-1}$ for the primary beam integrated spectrum, with Gaussian central velocities at $7924 \pm 12 \, \rm km\,s^{-1}$ and $7945 \pm 18 \, \rm km\,s^{-1}$, respectively. The systemic velocity derived from the $^{12}$CO(3--2) emission in the primary beam is thus consistent with that determined from the $^{12}$CO(1--0) and HI lines (Table~\ref{Mrk590properties}).

\subsection{Modeling the $^{12}$CO(3--2) Gas Kinematics}\label{kinematicmodels} 

Our preliminary modeling shows that the velocity field of the inner gas ring-like structure in the $^{12}$CO(3--2) maps is consistent with a simple rotating disk. We confirmed this using the Kinematic Molecular Simulation software of \citet{davisetal13}. The observed velocity map (Figure~\ref{momentmaps}b) is consistent with a rotating annulus of radius (to midpoint) of 0.93$\pm$0.01\,kpc and a width of 65\,pc with an inclination angle of 64$\substack{+3\\-1}$ degrees to our line of sight and a mean rotational velocity of 117\,km\,s$^{-1}$. Given the crude nature of this model, we do not graphically display the results or the residuals. 

We emphasize that this rotational model is not well constrained by the $^{12}$CO(3--2) data alone, but we quote the best fit solution of numerous trial models. All trials confirm that region C (Figure~\ref{spectregions}) contains a velocity component displaying too large a velocity deviation (by up to 80\,km\,s$^{-1}$) to be an integrated part of a rotating annulus. One reason for the non-unique solution is that the morphology of the emission displays an incomplete ring, perhaps due to lower amounts of CO gas in parts of the structure. However, the morphology and kinematics of the inner $^{12}$CO(3--2) gas structure is also consistent with an expanding shell (radial velocity $\sim$115\,km\,s$^{-1}$) with the gas component in region C unrelated thereto. The main limitation of this type of analysis is, however, insufficient information about the central gravitational potential of the galaxy. Our $^{12}$CO(3--2) line maps are at present the only dataset providing spatially resolved gas kinematics in the central kpc regions of Mrk\,590. We are attempting to obtain more data (e.g., IFU spectroscopy with the Very Large Telescope (VLT), and high resolution HI line maps with MERLIN) to better model the rotation curves and dynamics of the gas in the central kpc. These will enable us to better study signatures of non-rotational gas components, e.g., in/outflowing gas, and determine which (if any) physical processes are responsible for driving the gas inwards to fuel the AGN. 

\subsection{Molecular Gas Masses Derived from $^{12}$CO(3--2) Emission}\label{gasmass}

Here, we determine the molecular (H$_2$) gas mass in the central regions of Mrk\,590 in order to assess if it has significantly lower amounts of gas relative to other nearby Seyfert galaxies at hundreds of parsec and kpc scales. 

We do not detect any $^{12}$CO(3--2) emission down to the image sensitivity limits of 1$\sigma \sim 0.5{\rm\,mJy\,beam^{-1}}$ in the central 150\,pc, where the continuum emission is detected and the AGN is located. To confirm this, we examined the spectra of the inner 150\,pc using the datacubes without continuum subtraction, to ensure that there were no broad $^{12}$CO(3--2) components that were unintentionally removed during continuum subtraction. 

To derive an upper limit to the molecular gas mass in the central 150\,pc, we first estimate the upper limit of the velocity integrated line flux of $^{12}$CO(3--2) in this region. Assuming a signal smoothed over the expected FWHM velocity width of the emission line, $\Delta v_{\rm FWHM}$, the $3\sigma$ upper limit of the integrated flux can be estimated to be:
\begin{equation}\label{uplimit} 
I \leq 3\sigma \Delta v_{\rm FWHM} \sqrt{\dfrac{\Delta v}{\Delta v_{\rm FWHM}}} 
\end{equation}
where $\Delta v$ is the channel width of the spectrum (i.e., spectral resolution) at which the rms intensity, $\sigma$, is determined. The factor $\sqrt{{\Delta v}/{\Delta v_{\rm FWHM}}}$ accounts for the expected decrease in $\sigma$ for a channel width equivalent to $\Delta v_{\rm FWHM}$ \citep{wrobelwalker99}. In the inner 150\,pc, we assume that $\Delta v_{\rm FWHM} = 230\,\rm km\,s^{-1}$, similar to the FWHM of the integrated $^{12}$CO(3--2) spectra in the central 4$^{\prime\prime}$. For $\sigma = 0.5{\rm\,mJy\,beam^{-1}}$ and channel widths of $\Delta v = 20\,{\rm km\,s^{-1}}$, we estimate the upper limit of the integrated flux density of $^{12}$CO(3--2) in the central 150\,pc to be $I _{\rm CO(3-2)} \leq 0.1\,{\rm  Jy\,km\,s^{-1}}$. From this value of $I _{\rm CO(3-2)}$, the associated line luminosity is estimated using the equation \citep{solomonvandenbout05}:
\begin{align}\label{linelum} 
L'_{\rm CO(3-2)} = \,\,& 3.25 \times 10^7 \times \left( \dfrac{I_{\rm CO(3-2)}}{\rm Jy\,km\,s^{-1}} \right) \left( \dfrac{D_{\rm L}^2}{\rm Mpc} \right) \notag \\ 
&\times \left( \dfrac{\nu_{\rm rest}}{\rm GHz} \right)^{-2} (1+z)^{-1}\,{\rm K\,km\,s^{-1}\,pc^2} 
\end{align}
where $D_{\rm L}$ is the luminosity distance and $\nu_{\rm rest}$ is the rest frequency of the $^{12}$CO(3--2) line ($345.7959$\,GHz). For a given value of $\alpha_{\rm CO}$ (the H$_2$ mass to $^{12}$CO(1--0) line luminosity ratio), the H$_2$ gas mass is estimated as:
\begin{equation}\label{COmasscon} 
M({\rm H_2}) = \alpha_{\rm CO} L'_{\rm CO(1-0)} = \dfrac{\alpha_{\rm CO} L'_{\rm CO(3-2)} }{L'_{\rm CO(3-2)}/L'_{\rm CO(1-0)}} \,.
\end{equation}
We thus estimate the upper limit of the H$_2$ gas mass in the central 150\,pc to be:
\begin{align}\label{COmass} 
M({\rm H_2}) \leq \,\,& 5.3 \times 10^6\, {M_{\odot}} \left( \dfrac{\alpha_{\rm CO}}{4\,{M_{\odot}\rm \,(K \,km\,s^{-1}\,pc^{2}})^{-1}} \right) \notag \\ 
& \times \left( \dfrac{0.27}{L'_{\rm CO(3-2)}/L'_{\rm CO(1-0)}} \right)
\end{align}
where we have used the fiducial values of $\alpha_{\rm CO} = 4\,{M_{\odot}\rm \,(K \,km\,s^{-1}\,pc^{2}})^{-1}$, typical of giant molecular clouds in the disk of the Milky Way \citep{bolattoetal13}. We have also adopted a line luminosity ratio of $L'_{\rm CO(3-2)}/L'_{\rm CO(1-0)} = 0.27$ observed for the Milky Way \citep{carilliwalter13}. This value of $5.3 \times 10^6\, {M_{\odot}}$ is a very conservative upper limit of $M(\rm H_2)$. We note also that $M({\rm H_2}) \propto \sqrt{\Delta v_{\rm FWHM}}$, such that the estimated gas mass limits can be higher if the line widths are larger than assumed.

Stronger limits can be obtained if one were to adopt more realistic values of $\alpha_{\rm CO}$ and $L'_{\rm CO(3-2)}/L'_{\rm CO(1-0)}$, since it is unlikely that the gas conditions in the inner 150\,pc are similar to gas conditions averaged over the entire Milky Way. Since our upper limit on $L'_{\rm CO(2-1)}$ in Mrk\,590 is unable to provide any useful constraint on the excitation levels of the CO gas (described further in Section~\ref{CO21lim} below), we make use of values derived from other studies. \citet{sandstrometal13} find that in the central kpc regions of nearby galaxies, $\alpha_{\rm CO}$ can have much lower values of $\sim 0.8\,{M_{\odot}\,\rm (K \,km\,s^{-1}\,pc^{2}})^{-1}$ as typically found in ultraluminous infrared galaxies (ULIRGs). \citet{pereirasantaellaetal13} also find that the mean shape of the CO Spectral Line Energy Distribution (SLED, which describes the excitation levels of the CO gas and the relative line fluxes between the different CO transitions) for six local Seyfert galaxies resembles that of the two canonical starburst galaxies M82 and Arp\,220. In these starburst galaxies, $L'_{\rm CO(3-2)}/L'_{\rm CO(1-0)} \sim 0.9$ as also seen in quasars \citep{carilliwalter13}. Assuming these values of $\alpha_{\rm CO}$ and $L'_{\rm CO(3-2)}/L'_{\rm CO(1-0)}$, we obtain a stronger upper limit of $M({\rm H_2}) \lesssim 1.6 \times 10^5\, {M_{\odot}}$ in the central 150\,pc of Mrk\,590.

From the spatially integrated spectra of the central 4$^{\prime\prime}$ ($\sim$2\,kpc, Figure~\ref{spectotal}, top panel), we obtain an integrated flux density of 3.33 $\rm Jy\,km\,s^{-1}$ over all channels with CO line emission, giving an equivalent H$_2$ gas mass of $M({\rm H_2}) \lesssim 1.7 \times 10^8 \,M_{\odot}$. This assumes Milky Way values of $\alpha_{\rm CO} = 4\,{M_{\odot}\,\rm (K \,km\,s^{-1}\,pc^{2}})^{-1}$ and $L'_{\rm CO(3-2)}/L'_{\rm CO(1-0)} = 0.27$. This estimate is consistent with the typical gas masses of $10^7\,M_{\odot}$ to $10^{10}\,M_{\odot}$ found in the inner kpc regions of other nearby Seyfert galaxies and LINERs from the NUGA survey \citep[e.g.,][]{garcia-burilloetal03,combesetal04,combesetal09,kripsetal05,booneetal07,casasolaetal08} and in NGC\,4151 \citep{dumasetal10}. We note that these studies typically also adopt the Milky Way value of $\alpha_{\rm CO}$, so the comparisons are like for like. However, if $\alpha_{\rm CO} = 0.8\,{M_{\odot}\,\rm (K \,km\,s^{-1}\,pc^{2}})^{-1}$ in the circumnuclear regions and $L'_{\rm CO(3-2)}/L'_{\rm CO(1-0)} = 0.9$, the H$_2$ gas mass in the inner 4$''$ may be lower, i.e., $M({\rm H_2}) \sim 1.0 \times 10^7 \,M_{\odot}$. While the actual gas mass may lie anywhere between the higher and lower estimates, we adopt the lower value of $M({\rm H_2})$ for the rest of our discussions unless stated otherwise, since it appears more appropriate as discussed above. 

Integrating over all channels with $^{12}$CO(3--2) emission in the spatially integrated spectra of the 18$''$ primary beam, we obtain a total velocity-integrated flux density of 8.34 $\rm Jy\,km\,s^{-1}$. This gives us a total H$_2$ gas mass of $M({\rm H_2}) \sim 4.4 \times 10^8 \,M_{\odot}$, assuming the more appropriate Milky Way values of $\alpha_{\rm CO}$ and $L'_{\rm CO(3-2)}/L'_{\rm CO(1-0)}$ given the larger spatial scales of the emission. Based on the single dish $^{12}$CO(1--0) integrated flux densities \citep{maiolinoetal97}, we obtain $M({\rm H_2}) \sim 7.9 \times 10^9 \,M_{\odot}$ within the 55$''$ beam, for the cosmology and distance adopted in this current study. Our mass estimates from the ALMA observations are thus a factor of 18 lower than that estimated from the single dish observations, due to the smaller ALMA primary beam size. There is also very likely to be missing flux not recovered in our interferometric images due to spatial filtering of the extended emission. 

The values of $I _{\rm CO(3-2)}$, $L'_{\rm CO(3-2)}$ and $M({\rm H_2})$ in the inner 150\,pc and 2\,kpc are summarized in Table~\ref{resultssummary}. We discuss the implications of these results in Section~\ref{reservoirs}.

\begin{table}
\caption{Emission Line Properties, Gas Masses and Star Formation Rate Estimates in Mrk\,590.}
\label{resultssummary}
\begin{tabular}{lccc}
\hline
\hline
Property$^a$ & Units & Central & Central \\
 & & $\rm 180\,pc \times 120\,pc$ & 2\,kpc\\
\hline
$I_{\rm CO(3-2)}$ & (Jy\,km\,s$^{-1}$) 	&   $\leq$ 0.1 		&  3.33 \\
$L'_{\rm CO(3-2)}$ & (${\rm K\,km\,s^{-1}\,pc^2}$) & $\leq 3.6 \times 10^5 $ & $4.4 \times 10^7$\\
${M({\rm H_2})}^b$ & ($M_{\odot}$) & $\leq 1.6 \times 10^5$  & $1.0 \times 10^7$ \\
 & & [$\leq 5.3 \times 10^6$] & [$1.7 \times 10^8$] \\
\hline
$I_{\rm HCO(4-3)}$ & (Jy\,km\,s$^{-1}$) & ... & $\leq 0.48$ \\
$L'_{\rm HCO(4-3)}$ & (${\rm K\,km\,s^{-1}\,pc^2}$) & ... & $\leq1.5 \times 10^7$\\
$R^{\rm HCO43}_{\rm CO32}$ & ... & ... & $\leq 0.14$\\
\hline
$I_{\rm CO(2-1)}$ & (Jy\,km\,s$^{-1}$) 	& ...	& $\leq$ 1.6 \\
$L'_{\rm CO(2-1)}$ &  (${\rm K\,km\,s^{-1}\,pc^2}$) & ... & $\leq 1.3 \times 10^7$\\
\hline
SFR($\rm H_2$)$^b$ & ($M_{\odot}\,\rm yr^{-1}$) & $\leq 2 \times 10^{-4}$ & 0.004 \\
 & & [$\leq 0.009$] & [0.2] \\
${\rm \Sigma_{SFR}{(H_2)}}^b$ & ($M_{\odot}\, \rm yr^{-1} \, kpc^{-2}$)& $\leq 0.004$ & 0.0011\\
 & & [$\leq 0.2$] & [0.06]\\
SFR(H$\alpha$)$^c$ & ($M_{\odot}\,\rm yr^{-1}$) & ... & $\leq 0.15$\\
$\rm \Sigma_{SFR}(H\alpha)$ & ($M_{\odot}\,\rm yr^{-1}$) & ... & $\leq 0.13$\\
\hline
\end{tabular}
\begin{flushleft}
%$^a$SFR($\rm H_2$) and $\Sigma_{\rm {SFR(H2)}}$ are the SFR and SFR surface density derived based on the Kennicutt-Schmidt relation, using $M_{\rm H_2}$. 
%$^b$SFR(H$\alpha$) and $\Sigma_{\rm SFR(H\alpha)}$ are the SFR and SFR surface density derived from the H$\alpha$ luminosity. 
$^a I_{\rm X}$ is the velocity integrated flux density of emission line `$\rm X$'; $L'_{\rm X}$ is the luminosity of emission line `$\rm X$'; $M({\rm H_2})$ is the H$_2$ gas mass estimate; $R^{\rm HCO43}_{\rm CO32}$ is the ratio of the HCO$^+$(4--3) to $^{12}$CO(3--2) integrated flux densities. $\rm \Sigma_{SFR}$ is the star formation rate (SFR) surface density derived from either $M({\rm H_2})$ using the Kennicutt-Schmidt relation  (Section~\ref{SFRCO}) or the H$\alpha$ line luminosity (Section~\ref{SFRHalpha}), as indicated. \\
$^b$Estimates are based on $L'_{\rm CO(3-2)}/L'_{\rm CO(1-0)} \sim 0.9$ and $\alpha_{\rm CO} \sim 0.8\,{M_{\odot}\,\rm (K \,km\,s^{-1}\,pc^{2}})^{-1}$; the values in square brackets show more conservative estimates based on Milky Way values of $L'_{\rm CO(3-2)}/L'_{\rm CO(1-0)} \sim 0.27$ and $\alpha_{\rm CO} \sim 4\,{M_{\odot}\,\rm (K \,km\,s^{-1}\,pc^{2}})^{-1}$. \\
$^c$Measured in a rectangular region of size $4'' \times 1.2''$ ($\sim$2\,kpc $\times$ 0.6\,kpc).
\end{flushleft}
\end{table}  

\subsection{Upper Limits on HCO$^+$(4--3) Emission}\label{HCOlim}

The critical density for exciting HCO$^+$ is $6.5 \times 10^{6}$~cm$^{-3}$, as opposed to $3.6 \times 10^{4}$~cm$^{-3}$ for $^{12}$CO(3--2) \citep{carilliwalter13}. Along with HCN, HCO$^+$ emission is therefore a good tracer of dense gas distributions in the centre of local AGNs \citep[e.g., for NGC\,1097,][]{izumietal13,martinetal15,onishietal15}, close to the black hole. The relative abundances of these dense gas tracers with respect to CO are strong indicators of the presence of starbursts \citep{gaosolomon04}, which in turn have been found to be connected to AGN activity (Section~\ref{SFR}). We therefore examine the upper limits of the HCO$^+$(4--3) emission line flux in Mrk\,590, to determine if its relative abundance to CO is comparable to that of other local AGNs. 
	
In the ALMA HCO$^+$(4--3) spectra integrated spatially over the central 4$^{\prime\prime}$ (Figure~\ref{spectotal}, top panel), encompassing the inner gas ring, we do not detect any line emission down to a $1\sigma$ rms of 2.3 mJy. To place an upper limit on the  integrated flux density of HCO$^+$(4--3) in this region, we assume the HCO$^+$(4--3) gas traces the $^{12}$CO(3--2) line emitting regions and adopt a line width of 230\,km\,s$^{-1}$, equivalent to the FWHM of the $^{12}$CO(3--2) line in the central 4$^{\prime\prime}$.  Following Equation~\ref{uplimit}, we constrain the velocity-integrated flux density of HCO$^+$(4--3) to $I_{\rm HCO(4-3)} \lesssim \rm 0.48\, Jy\,km\,s^{-1}$ within the central 4$''$, and thus obtain $L'_{\rm HCO(4-3)} \lesssim 1.5 \times 10^7\, {\rm K\,km\,s^{-1}\,pc^2}$. 

For the spectra integrated over the entire 18$^{\prime\prime}$ ALMA primary beam (Figure~\ref{spectotal}, bottom panel), the HCO$^+$(4--3) line is also undetected down to a $1\sigma$ rms of 4.5\,mJy. This yields $I_{\rm HCO(4-3)} \lesssim \rm 0.84 \, Jy\,km\,s^{-1}$, in turn giving $L'_{\rm HCO(4-3)} \lesssim 2.6 \times 10^7 \,{\rm K\,km\,s^{-1}\,pc^2}$. This assumes a line width of 190\,km\,s$^{-1}$, similar to the FWHM of the $^{12}$CO(3--2) line over the 18$^{\prime\prime}$ primary beam. This upper limit of $L'_{\rm HCO(4-3)}$ is consistent with the typical values of $L'_{\rm HCO(4-3)} \sim 10^{6}\, {\rm K\,km\,s^{-1}\,pc^2}$ observed in nearby active galaxies with the single-dish Atacama Pathfinder EXperiment (APEX) telescope \citep{zhangetal14}. %\citet{zhangetal14} also find that starbursting galaxies typically have 

As a measure of the relative abundance of HCO$^+$ to CO, we make use of the ratio of the integrated flux densities, $R^{\rm HCO43}_{\rm CO32} = I_{\rm HCO(4-3)}/I_{\rm CO(3-2)}$. We obtain upper limits of  $R^{\rm HCO43}_{\rm CO32} \lesssim 0.14$ and $R^{\rm HCO43}_{\rm CO32} \lesssim 0.1$ for the central 4$^{\prime\prime}$ and the full primary beam, respectively. This result is consistent with other local AGNs, such as NGC\,1433 where $R^{\rm HCO43}_{\rm CO32} \lesssim 1/60$ at 3$\sigma$ \citep{combesetal13}. Even when HCO$^+$(4--3) has been detected in nearby Seyfert nuclei, the line is typically $\sim$10 to 100 times weaker than that of $^{12}$CO(3--2) \citep[e.g.,][]{casasolaetal11,combesetal14,garcia-burilloetal14}. Since our constraint of $R^{\rm HCO43}_{\rm CO32} \lesssim 0.1$ is weak compared to these other studies, we are unable to make meaningful comparisons of the dense gas fraction in Mrk\,590 with that of other local Seyferts. We note however, that our results are consistent with Mrk\,590 being a typical Seyfert galaxy on kpc scales.

\subsection{Upper Limits on $^{12}$CO(2--1) Emission}\label{CO21lim}

We derive upper limits on the integrated $^{12}$CO(2--1) line flux from the SMA spectra, to obtain constraints on the excitation levels of the CO gas. This in turn is important in the estimation of H$_2$ gas masses from higher excitation ($J > 1$) CO lines (Section~\ref{gasmass}). Integrating the spectra from the SMA upper side-band over the inner 4$''$ and 18$''$, corresponding to the inner gas ring and the ALMA primary beam, we obtain an rms flux density of 7.2 mJy and 27.1 mJy, respectively, for channel widths of 25\,km\,s$^{-1}$. The velocity-integrated flux densities of the $^{12}$CO(2--1) line are thus constrained to be $I_{\rm CO(2-1)}\lesssim \rm 1.6\, Jy\,km\,s^{-1}$ within the inner 4$''$ and $I_{\rm CO(2-1)} \lesssim \rm 5.6\, Jy\,km\,s^{-1}$ in the inner 18$''$. Again, we assume that the $^{12}$CO(2--1) emission originates from the same regions traced by $^{12}$CO(3--2). Therefore, the $^{12}$CO(2--1) FWHM widths are assumed to be 230\,km\,s$^{-1}$ and 190\,km\,s$^{-1}$ respectively, corresponding to the $^{12}$CO(3--2) line widths in these respective regions. These upper limits are less than a factor of two lower than the integrated flux densities of $^{12}$CO(3--2) encompassing the same regions in the ALMA data. Generally, $^{12}$CO(2--1) fluxes are a factor of two lower than that of $^{12}$CO(3--2) in quasars, and can be even lower for submillimeter galaxies and star forming galaxies \citep{carilliwalter13}. It is therefore not surprising that we detect no significant emission of $^{12}$CO(2--1) in the SMA spectra. We are also unable to obtain any strong constraints on the shape of the CO SLED for Mrk\,590. We have secured ALMA Cycle\,3 observing time to try to detect the $^{12}$CO(1--0) and $^{12}$CO(6--5) lines to better constrain the CO SLED, thereby reducing the uncertainties in the estimates of the molecular gas masses. 

\subsection{Circumnuclear Star Formation Rates}\label{SFR}

In hydrodynamical simulations of gas inflow from 10\,kpc galactic scales down to 0.1\,pc scales to fuel an AGN, nuclear star formation rates (SFRs) are found to correlate with black hole mass accretion rates \citep{hopkinsquataert10}. Since the nuclear SFR may provide important clues on the nuclear fueling of Mrk\,590, and stellar outflows provide a viable mechanism for transporting the gas into the centre to feed the black hole, we estimate the SFR in the circumnuclear regions. We derive the SFR using two independent methods, (1) based on the Kennicutt-Schmidt relation using $M({\rm H_2})$ (or its upper limit) derived from the $^{12}$CO(3--2) emission (Section~\ref{gasmass}) and (2) from the H$\alpha$ luminosity $L_{\rm H\alpha}$. The 344\,GHz and 219\,GHz continuum emission, as well as the 1.4\,GHz radio continuum emission, provide alternative means of estimating the circumnuclear SFR. However, it is very likely that the continuum emission at these wavelengths contains a significant AGN contribution to the flux, such that the far-IR and radio based SFR estimates will be biased high, providing only upper limits.

\subsubsection{SFR Estimates Derived from the H$_2$ Gas Mass}\label{SFRCO}

Based on the upper limit of $M(\rm H_2) \lesssim 1.6 \times 10^5\, {M_{\odot}}$, derived from the $^{12}$CO(3--2) observations (Section~\ref{gasmass}), we estimate the SFR in the central 150\,pc to be ${\rm SFR(H_2)} \lesssim 2 \times 10^{-4}\,M_{\odot}\, \rm yr^{-1}$ following the Kennicutt-Schmidt relation \citep{kennicutt98}. The corresponding SFR(H$_2$) surface density is ${\rm \Sigma_{SFR}{(H_2)}}\lesssim 0.004 \,M_{\odot}\, \rm yr^{-1} \, kpc^{-2}$. We note that our SFR limits assume negligible HI gas masses in the centre. This is a reasonable assumption, since it is very likely that the gas phase is predominantly molecular in the centre, as seen in other local AGNs \citep{walteretal08, haanetal08, bigieletal08}. For example, the mean H$_2$ gas mass surface densities are found to be at least 20--40 times higher than that of atomic gas in the NUGA sample of AGNs \citep{casasolaetal15}. Additionally, the gas surface densities of nearby AGNs are in a regime in which the model by \citet{krumholtzetal09} shows that H$_2$ is expected to dominate \citep[an example is shown in Fig. 8 of][]{casasolaetal15}. The inferred SFR limit for Mrk\,590 is in the lower range of SFRs observed in the central $\sim 65$\,pc of 29 nearby Seyfert galaxies \citep{esquejetal14}, derived from the nuclear 11.3\,$\mu$m PAH feature after subtracting AGN contributions. \citet{esquejetal14} obtain SFR estimates of ${\rm SFR(PAH)} \sim 0.01\,M_{\odot}\, \rm yr^{-1}$ to $1.2 \,M_{\odot}\, {\rm yr^{-1}}$ for sources with nuclear 11.3\,$\mu$m PAH detections, and ${\rm SFR(PAH)} \lesssim 0.01\, M_{\odot}\, {\rm yr^{-1}}$ to $0.2\, M_{\odot}\, \rm yr^{-1}$ for sources with non-detections. Even assuming the more conservative upper limit of $M({\rm H_2}) \lesssim 5.3 \times 10^6\, {M_{\odot}}$, the upper limit on the total SFR within the central 150\,pc is still very low, with ${\rm SFR(H_2)} \lesssim 0.009\,M_{\odot}\, \rm yr^{-1}$. However, a higher, less stringent limit of ${\rm \Sigma_{SFR}{(H_2)}} \lesssim 0.2 \,M_{\odot}\, \rm yr^{-1} \, kpc^{-2}$ is obtained.

In the inner 4$''$ (2\,kpc), we estimate ${\rm SFR(H_2)} \sim 0.2 \,M_{\odot}\, \rm yr^{-1}$ and ${\rm \Sigma_{SFR}{(H_2)}} \sim 0.06 \,M_{\odot}\, \rm yr^{-1} \, kpc^{-2}$ from the inferred $M(\rm H_2) \sim 1.7 \times 10^8\, {M_{\odot}}$, assuming Milky Way values of $\alpha_{\rm CO}$ and $L'_{\rm CO(3-2)}/L'_{\rm CO(1-0)}$. If we adopt starburst galaxy-like conditions and use the lower value of $M({\rm H_2}) \sim 1 \times 10^7\, {M_{\odot}}$, we estimate ${\rm SFR(H_2)} \sim 0.004 \,M_{\odot}\, \rm yr^{-1}$ and ${\rm \Sigma_{SFR}{(H_2)}} \sim 0.0011 \,M_{\odot}\, \rm yr^{-1} \, kpc^{-2}$. As with the estimates of $M({\rm H_2})$, the actual value of the SFR may be somewhere in between the range bordered by the lower and higher estimates. 

The SFRs and H$_2$ gas masses in the central 60\,pc of nearby Seyfert nuclei are found to be consistent with the Kennicutt-Schmidt relation \citep{hicksetal09}. \citet{casasolaetal15} also find no degradation of this  relation down to 20\,pc scales in four nearby low-luminosity AGNs from the NUGA sample. We thus expect the SFR(H$_2$) estimates to be representative. The main source of error in the estimates is the uncertainty in $\alpha_{\rm CO}$ used in estimating the H$_2$ gas masses, since $\Sigma_{\rm SFR} \propto (\Sigma_{\rm gas})^{1.4}$, where $\Sigma_{\rm gas}$ is the gas mass surface density. We summarize in Table~\ref{resultssummary} the values of the SFRs and $\Sigma_{\rm SFR}$ estimated for the inner 150\,pc and inner 2\,kpc of Mrk\,590, for each case listing both values derived based on the different assumptions about the gas conditions. 

\subsubsection{SFR Estimate Derived from the H$\alpha$ Luminosity}\label{SFRHalpha}

As an independent check of the SFR(H$_2$) estimates, we derive also the SFR from the H$\alpha$ emission line, SFR(H$\alpha$). We use the narrow component of the H$\alpha$ line, as the broad component is not representative of the star-forming regions. The narrow H$\alpha$ line has a flux of $1.2 \times 10^{-14 }\,\rm erg\,s^{-1}\,cm^{-2}$ \citep{denneyetal14}, from the spectrum of Mrk\,590 obtained in 2013 with the Large Binocular Telescope. This yields a H$\alpha$ luminosity, $L_{\rm H\alpha} \sim 1.9 \times 10^{40}\, \rm ergs \, s^{-1}$, from which we infer ${\rm SFR(H\alpha)} \sim 0.15 \,M_{\odot}\, \rm yr^{-1}$ following \citet{kennicutt98}. While the present value of $L_{\rm H\alpha}$ is very low in comparison to its value in the 1990s when the AGN was accreting at a much higher rate, there is still likely to be an AGN contribution to $L_{\rm H\alpha}$, as evidenced by the presence of a weak broad H$\alpha$ component \citep{denneyetal14}. The AGN therefore has not fully turned off and hence the estimated SFR(H$\alpha$) is likely just an upper limit.
 
The SFR(H$\alpha$) value is much higher than SFR(H$_2$) estimated based on our CO observations in the central 150\,pc. This is to be expected since $L_{\rm H\alpha}$ is extracted from a rectangular region of $4'' \times 1.2''$ ($\sim$2\,kpc $\times$ 0.6\,kpc). The SFR surface density derived from $L_{\rm H\alpha}$ is  $\Sigma_{\rm SFR}({\rm H\alpha}) = 0.13\,M_{\odot}\, \rm yr^{-1} \, kpc^{-2}$, consistent with the upper limit of $\rm \Sigma_{SFR}{(H_2)}$ derived for the central 150\,pc. SFR(H$\alpha$) is also consistent with the estimated $\rm SFR{(H_2)}$ in the central 4$''$, to within the uncertainties arising from the use of different conversion factors, $\alpha_{\rm CO}$, in deriving the latter (Table~\ref{resultssummary}).  

\section{Discussion}\label{discussion} 

\subsection{Circumnuclear Gas Reservoirs and Distributions: How Unique is Mrk\,590?}\label{reservoirs}

As we detect no CO emission in the central 180\,pc $\times$ 120\,pc, assuming $L'_{\rm CO(3-2)}/L'_{\rm CO(1-0)} \sim 0.9$ and $\alpha_{\rm CO} \sim 0.8\,{M_{\odot}\,\rm (K \,km\,s^{-1}\,pc^{2}})^{-1}$ we estimate the molecular gas mass in the central $\sim$150\,pc to be $M({\rm H_2})\lesssim 1.6 \times 10^5\, {M_{\odot}}$ (Section~\ref{gasmass}). This is no more than that of a typical giant molecular gas cloud in the Milky Way Galaxy \citep[$\sim 10^4\,M_{\odot}$ to $10^6\,M_{\odot}$;][]{solomonetal87}. We observe a ring-like structure of CO gas containing $M({\rm H_2}) \sim 10^7\,M_{\odot}$ of molecular gas in the central 2\,kpc, where a gas clump with disturbed kinematics located just 200\,pc west of the AGN may be fueling the centre. 

The CO gas morphology of Mrk\,590, however, is not unique among nearby Seyfert galaxies. A similar central structure and CO distribution has been observed for the Seyfert 1.5 galaxy NGC\,4151 \citep{dumasetal10}, for example. \citet{dumasetal10} detect no CO emission in the inner 150\,pc of NGC\,4151, with the H$_2$ gas mass also constrained to less than $10^5\, M_{\odot}$. Two gas lanes of kpc-length scale form a partial ring at a distance of 1\,kpc from the centre of NGC\,4151. \citet{dumasetal10} also observe a gas clump located 3$''$ ($\sim$200\,pc) away from the centre with non-rotational kinematics that appears to be fueling the centre. It is intriguing that NGC\,4151 is also a changing-look AGN (Section~\ref{introduction}). 

On kpc scales, Mrk\,590 does not have significantly less gas than other local Seyfert galaxies. The H$_2$ gas mass estimates of Mrk\,590 at scales of a kpc to tens of kpc, as determined from our $^{12}$CO(3--2) data and by \citet{maiolinoetal97}, do not show significant deviation from that of other nearby Seyfert galaxies (Section~\ref{gasmass}). Notably, the H${_2}$ gas mass of $9.5 \times 10^9\,M_{\odot}$ in Mrk\,590 derived from the $^{12}$CO(1--0) line by \citet{maiolinoetal97} is just slightly above the peak of the H${_2}$ mass distribution ($\sim 2 \times 10^9\,M_{\odot}$) of their entire sample of 94 nearby Seyfert galaxies. The fraction of dense gas and excitation states of the CO gas in Mrk\,590, as far as can be constrained from the non-detection of the HCO$^+$(4--3) and $^{12}$CO(2--1) emission lines (Sections~\ref{HCOlim} and \ref{CO21lim}), are also consistent with those of other local AGN host galaxies. At these spatial scales, there is nothing peculiar that sets Mrk\,590 apart. Any peculiarities would have to exist within the central 100\,pc or most likely within the central 1\,pc to 10\,pc, unreachable by our present observations, as one would expect based on the observed variability timescales of decades and light-travel time arguments. 

On a related note, observations of gas reservoirs and transport processes at kpc scales and even hundreds of pc scales may provide no direct indication of ongoing AGN activity, accretion strengths, or whether an AGN is turned on or off. For a sample of 95 X-ray selected AGNs from the \textit{Chandra} COSMOS survey, \citet{cisternasetal15} find that the presence of large-scale stellar bars do not correlate with AGN accretion strengths, and find no significant differences in the bar fractions of AGN hosts and normal quiescent spirals. \citet{cheungetal15} also find no significant difference in the fraction of AGNs located in barred or non-barred galaxies up to a redshift of 1. These recent studies confirm that the occurrence and efficiency of AGN fueling is independent of the structure of the galaxy on large scales, confirming the results of earlier studies \citep[e.g.,][]{hoetal97,mulchaeyregan97}. Even on smaller scales of a few hundred parsecs, the presence of nuclear bars do not correlate with AGN activity, as \citet{martinipogge99} find only five galaxies with nuclear bars among their sample of 24 Seyfert galaxies. While bars do indeed play a key role in driving gas from the galaxy disk into the central kpc regions, other mechanisms operating at smaller scales are required to drive the gas into the central tens of pc on timescales more relevant to the currently observed rate of accretion in an AGN. In fact, the NUGA program \citep[e.g.,][]{garcia-burilloetal03} find the gas within the central kpc regions to be characterized by a wide variety of gas morphologies, revealing that no single transport mechanism is responsible for driving the gas inwards to fuel the AGN. As mentioned in Section~\ref{introduction}, not all AGNs in the NUGA sample show evidence of fueling at hundreds of pc scales. Therefore, the morphology of the gas and the presence (or absence) of gas transport mechanisms operating at hundreds of pc scales also do not correlate with current AGN activity. Mrk\,590 is direct evidence of this, considering that the central engine turned off within a span of a few decades, while the 100\,pc and kpc scale structures will not evolve on these time-scales. Note however, that \citet{hicksetal13} have shown for their matched sample of 10 Seyfert and quiescent galaxies that within a 250\,pc radius, AGNs have a higher concentration of H$_2$ gas (and young stars) than quiescent galaxies. 

\subsection{Nuclear Fueling and Outflows: Is the AGN in Mrk\,590 Running out of Gas?}\label{dutycycles}

\subsubsection{Gas Consumption Time-scales}
 
Due to the high efficiencies of converting rest mass to energy in black hole accretion, AGNs do not require large amounts of gas to fuel the central supermassive black hole \citep[e.g.,][]{peterson97}. Even with the stronger H$_{2}$ gas mass limit of $1.6 \times 10^5\, {M_{\odot}}$ (Section~\ref{gasmass}), there is potentially sufficient gas to support the AGN activity in Mrk\,590 for $2.6 \times 10^5$ years at the Eddington accretion limit, and longer at lower rates of accretion; our Eddington mass accretion rate estimate is based on a black hole mass of $4.75 \times 10^7\,{M}_{\odot}$ measured from reverberation mapping \citep{petersonetal04} a 10\% mass-to-energy conversion efficiency \citep{marconietal04}, and solar-metallicity gas. Therefore, based on this gas mass limit, we are unable to rule out the possibility that Mrk\,590 may only be experiencing a temporary feeding break and may thus turn on again as observed for NGC\,4151.

\subsubsection{Gas Transport Mechanisms}

Are outflows from stars in the inner 150\,pc fueling the AGN in Mrk\,590? Based on the 2013 AGN bolometric luminosity estimated by \citet{denneyetal14} and an assumed 10\% mass-to-energy conversion efficiency, the AGN mass accretion rate at present is $6 \times 10^{-4} \,M_{\odot}\, \rm yr^{-1}$. This is a factor of three larger than the upper limit of the current ${\rm SFR(H_2)} \lesssim 2 \times 10^{-4}\, M_{\odot}\, \rm yr^{-1}$ (Section~\ref{SFR}) estimated for the central 150\,pc. At the present accretion rate, the AGN is still more efficient at consuming the gas in the inner 150\,pc compared to the star formation process, provided the gas can be transported into the central parsec to fuel the black hole. Even assuming optimistically that the stellar mass outflow rate is of the same order of magnitude as the SFR, as seen in galaxy-wide stellar outflows \citep[e.g.,][]{veilleuxetal05}, it is unlikely that outflows from stars in the inner 150\,pc are sufficient to continue fueling the central engine of Mrk\,590 at the present accretion rate. By no means could such stellar mass flows fuel the AGN in the 1990s when it was accreting at the higher rate of $0.1 \, M_{\odot}\, \rm yr^{-1}$ as indicated by its $L_{\rm bol}$ value (Table~\ref{Mrk590properties}). Unless the SFR and mass outflow rate were both much higher in the past, other gas transport mechanisms must be responsible for driving the gas inward on scales less than 100\,pc to fuel the AGN. This gas may be funneled down through gravitational instabilities, viscous torques, or dynamical friction of massive clouds against the old bulge stars \citep[e.g.,][]{combes02,jogee06}.

Why is there little detectable CO gas in the central 150\,pc? Circumnuclear gas rings at distances of $\sim$1\,kpc from the centre, such as those observed in Mrk\,590 and NGC\,4151 \citep{dumasetal10}, are often associated with Lindblad resonances, which can act as dynamical barriers that prevent further inflow of gas into the centre of the ring \citep[e.g.,][]{reganteuben04}. The elongated nuclear ring in the central 2\,kpc of Mrk\,590 may correspond to a resonance with the nuclear bar (if present, as discussed in Section~\ref{resultsCO}). Most of the circumnuclear gas would be maintained in the resonant ring by the nuclear bar torques, while a small amount of gas (below the threshold of detectability of our CO observations) could be transported into the centre through e.g., weak viscous torques, to trigger both the star formation and black hole accretion. To confirm this requires additional data constraining the dynamics (we will address this in future work). A continuous inflow rate of order $10^{-5} \,M_{\odot}\, \rm yr^{-1}$ is required to maintain the current AGN accretion rate and SFR in the central 150\,pc. However, it is possible that the gas inflow into the centre occurs in discrete episodes \citep[e.g., ][]{garcia-burilloetal05}, which then trigger episodic bursts of both nuclear star formation and AGN fueling. Since there is expected to be a time lag between the onset of nuclear star formation and the transport of gas into the central pc to fuel the AGN \citep[typically $10^7$ years on spatial scales of 10\,pc or less and $10^8$ years on scales of 1\,kpc,][]{hopkins12}, this may explain the current SFR being lower relative to the present AGN accretion rate in Mrk\,590. The burst of star formation may have preceded the nuclear accretion and hence began turning off earlier (if related). This scenario is consistent with the findings of the NUGA survey, where clear evidence of fueling at hundreds of pc scales is observed in only one third of the sample, suggesting that gas transport mechanisms on these scales are possibly activated only a third of the time in a particular galaxy.

The western clump (component C, Figure~\ref{spectregions}), with $M({\rm H_2}) \sim 1.2\times 10^{6}\,M_{\odot}$, is the most likely source of replenishment of gas into the central 150\,pc of Mrk\,590. This is evidenced by the component being located closest to the AGN in both configuration and velocity space (Figures~\ref{momentmaps} and \ref{pv}), as well as by the detection of disturbed kinematics (Section~\ref{kinematicmodels}). If indeed gas is transported into the central 150\,pc through this region, there is, in fact, enough gas in component C to fuel the AGN for a further $10^6$ years assuming Eddington-limited accretion. 

\subsubsection{No Detection of High-Velocity Molecular Outflows} 

Are outflows removing the central gas in Mrk\,590? High velocity molecular outflows have been discovered in nearby active galaxies \citep{ciconeetal14}. Although these kpc-scale outflows are predominantly due to stellar feedback, \citet{ciconeetal14} find evidence that AGNs may significantly boost the rate of mass outflows, based on observed correlations between mass outflow rates and AGN bolometric luminosities. Such molecular outflows may be entrained by radio jets \citep[e.g.,][]{ragacabrit93} or coupled to accretion disk winds. Indeed, \citet{tombesietal15} measure the energetics of accretion disk winds and large-scale molecular outflows in the quasar IRAS F11119+3257, and find that energy is conserved, consistent with the two processes being energetically coupled. 

\citet{guptaetal15} report the detection of X-ray absorbers with velocities of 0.176$c$ and 0.0738$c$ in Mrk\,590, indicating the presence of ultra-fast outflows at distances of $10^{-4}$\,pc from the central black hole, inside the broad-line region. However, it is not known if these high velocity winds are actually coupled to outflows of a significant gas mass at larger scales. Since we do not detect any $^{12}$CO(3--2) line emission in the central 150\,pc, we are unable to establish if molecular outflows may be depleting the gas at the very centre. The spatially integrated spectra in the inner 4$''$ and over the ALMA primary beam (Figure~\ref{spectotal}) do not show the presence of very broad $^{12}$CO(3--2) line components or wings (with velocities greater than the rotational velocities of the galaxy, i.e., $v \rm > 300\,km\,s^{-1}$) down to our sensitivity limits, that would be indicative of kpc-scale high velocity outflows. Therefore, it is unlikely that \textit{strong} molecular outflows are responsible for depleting the circumnuclear gas in Mrk\,590. Higher signal-to-noise spectra are needed to determine if there are wings or broad components of the $^{12}$CO(3--2) line that are too weak to be detected in our current dataset. The map of the extended $\rm [O\, \textsc{iii}]$ line emission of Mrk\,590 \citep{schmittetal03} also does not reveal biconical structures often associated with gas outflows in the narrow-line region \citep[e.g.,][]{crenshawetal10,fischeretal10,fischeretal13}. However, it is possible that the axis of the outflows are aligned towards our line-of-sight such that we do not observe the biconical structure. High-resolution IFU spectroscopy of the extended $\rm [O\, \textsc{iii}]$ line emission is needed to reveal possible signatures of outflows in the narrow-line region of this AGN.

\subsection{Disparity between ALMA Continuum and $^{12}$CO(3--2) Morphologies}\label{contdisparity}

Dust emission is expected to trace CO emission in star forming regions. As is clear from Figure~\ref{momentmaps}, we do not detect the dust continuum tracing the CO emission in the inner CO gas ring. Such disparities between the morphologies of the (sub-)mm continuum and CO line emission have also been observed in the Seyfert 1.5 galaxy NGC\,4151 \citep{dumasetal10} and the Seyfert\,2 galaxy NGC\,1433 \citep{combesetal13}. Since the intensity of dust emission is proportional to the dust temperature in the Rayleigh-Jeans domain, it is possible that the dust temperature in the inner gas ring is not high enough for the emission to be detected. Another possible explanation is that the extended continuum emission may have been filtered out spatially by the high angular resolution interferometric observations, as suggested by \cite{combesetal13}. This spatial filtering effect is typically worse in continuum images since the emission is spatially extended across all channels, whereas the line emission tends to be separated into smaller emitting regions in each frequency channel. 

Conversely, we detect the 344\,GHz continuum emission in the central 150\,pc where we do not detect any $^{12}$CO(3--2) emission. This suggests that there is warmer dust emission or other additional sources of continuum emission within the inner 150\,pc, which we discuss next.

\subsection{Origin of the 344\,GHz Continuum Emission in the Central 150\,pc}\label{contorigin}

The continuum emission detected by ALMA in the central 150\,pc can arise from thermal processes (e.g., free-free emission, dust heating by the central AGN and/or stars), non-thermal processes (e.g., synchrotron radiation from an AGN jet), or a combination of both. If this emission is dominated by dust heating, the derived dust mass provides an independent estimate of the total neutral (i.e., $\rm H_2 + HI$) gas mass in the centre. If this emission is AGN-dominated, understanding its origin and monitoring its possible variability will be important for studying the coupling of energetics between the component emitting at (sub-)mm wavelengths to that of the accretion disk, radio-emitting component, and broad-line region.

\subsubsection{Thermal Dust Emission}\label{thermaldust}
 
We first consider the scenario where the central continuum emission originates from thermal dust emission. \citet{garciaespinosa01} measure a continuum flux density of 2.59\,Jy for Mrk\,590 at 90\,$\mu$m using the \textit{Infrared Space Observatory (ISO)}. Combining this with our ALMA 344\,GHz continuum flux density, we estimate the spectral index to be $\alpha_{344\,{\rm GHz}}^{90\mu{\rm m}} \sim 3.56$. We note, however, that the \textit{ISO} observations are made with a larger beam size of order tens of arcseconds, and for that reason may contain far-IR emission from a much larger region than the continuum emission detected with ALMA. The spectral index calculated is thus an upper limit, and is consistent with thermal emission in the Rayleigh-Jeans domain (where $\alpha \sim 2$). 

If indeed thermal dust emission is the source of the continuum emission, we can gain a better understanding of whether this radiation originates from dust heated by the AGN or by stars in the central 150\,pc, by examining the near-IR to far-IR spectral energy distribution (SED). If well determined, the SED also allows us to derive the temperature and mass of the far-IR emitting dust component. We include in our analysis published measurements (summarised in Table~\ref{continuumfluxpub}) of the near-IR to far-IR flux densities of Mrk\,590 observed using the United Kingdom Infrared Telescope (UKIRT), the VLT VISIR instrument, \textit{ISO}, and the \textit{Infrared Astronomical Satellite (IRAS)}. We use the far-IR SED fitting code developed by \citet{casey12} to fit a near-IR power-law distribution and a modified blackbody (greybody) spectrum to the observed IR data points, along with our ALMA 344\,GHz flux density. For this model fit, we exclude the SMA continuum data point, as the SMA synthesized beam encompasses the molecular gas ring, and therefore likely includes additional thermal dust emission from the ring. Here we are interested only in the continuum emission in the central 150\,pc measured by ALMA. Although the UKIRT, \textit{ISO} and \textit{IRAS} flux densities are also measured with much lower angular resolutions compared to ALMA, the emission at these higher frequencies are more likely to be dominated by the AGN and thus by the continuum emission from within the inner 150\,pc. Further details on the model fitting and the associated caveats are outlined in Appendix~\ref{appendix}. Figure~\ref{sedfit} presents the resulting model SED fit (solid black curve) to the data.

\begin{table*}
\centering
\caption{Published Near-IR to Far-IR and Radio Continuum Flux Densities for Mrk\,590}
\label{continuumfluxpub}
\begin{tabular}{lccrrlc}
\hline
\hline
Instrument &  Date of & Angular & \multicolumn{1}{c}{Observing} &  \multicolumn{2}{c}{Flux Density} & Ref.  \\
& Observations & Resolution & \multicolumn{1}{c}{Wavelength} &  &  \\
 & (Year) & ($''$) & \multicolumn{1}{c}{($\mu$m)} & \multicolumn{2}{c}{(mJy)}& \\
 \hline
United Kingdom Infrared Telescope & 1982 & $\sim$5 & 1.2 & 16.8 & $\pm$\,\,\,1.3 &  1\\
(UKIRT) & &  & 1.7 & 23.1 & $\pm$\,\,\,0.7 &  \\
& &  & 2.2 & 26.6 & $\pm$\,\,\,0.8 &  \\
 & &  & 3.5 & 46.9 & $\pm$\,\,\,3.1 &  \\
 & &  & 4.8 & 64 & $\pm$\,\,\,13 &  \\
\hline
 Very Large Telescope & 2006 & $\sim$0.35 &  10.49 & 75.9 & $\pm$\,\,\,20.9 & 2\\
 (VLT) & & & 11.25 & 75.0 & $\pm$\,\,\,2.1 & \\
 & &  & 12.81 & 106.3 & $\pm$\,\,\,13.3 & \\
\hline
 \textit{Infrared Space Observatory} & 1995 -1998 & $\sim$1.5 - 90 & 16 & 460 & $\pm$\,\,\,138 & 3\\
(\textit{ISO}) & &  & 25 & 260 & $\pm$\,\,\,78 & \\
 &  &  & 60 & 3060 & $\pm$\,\,\,918 & \\
 &  &  & 90 & 2590 & $\pm$\,\,\,777 & \\
\hline
\textit{Infrared Astronomical Satellite} & 1983 & $\sim$240 & 60 & 530 & $\pm$\,\,\,90 & 4\\
 (\textit{IRAS}) &  & & 100 & 1410 & $\pm$\,\,\,300 & \\
\hline
Owens Valley Radio Observatory (OVRO) & 1983 & $\sim$90 & \multicolumn{1}{r}{$1.5\times 10^{4}$} & 3.6 & $\pm$\,\,\,0.6 & 5\\
Very Large Array (D-configuration) & 1983 & \,\,$\sim$90$^a$ & \multicolumn{1}{r}{$6\times 10^{4}$} & 7.6 & $\pm$\,\,\,0.4 & 5\\
Very Large Array (D-configuration) & 1983 & $\sim$90 & \multicolumn{1}{r}{$2\times 10^{5}$} & 11.2 & $\pm$\,\,\,1.4 & 5\\
Very Large Array (A-configuration) & 1991 & $\sim$0.3 & \multicolumn{1}{r}{$3.6\times 10^{4}$} & 3.53 & $\pm$\,\,\,0.2 & 6 \\
Multi-Element Radio Linked &  &  &  &  &  & \\
Interferometer Network (MERLIN)  & 1995 & $\sim$0.2 & \multicolumn{1}{r}{$2\times 10^{5}$} & 6.70 & $\pm$\,\,\,0.29 & 7\\
\hline
\end{tabular}
\begin{flushleft}
$^a$uv-tapering applied to image.\\
References. -- (1)~\citet{wardetal87}. (2)~\citet{horstetal09}. (3)~\citet{garciaespinosa01}. (4)~\citet{edelsonetal87}. (5)~\citet{edelson87}. (6)~\citet{kukulaetal95}. (7)~\citet{theanetal01}. 
\end{flushleft}
\end{table*}

\begin{figure}
\begin{center}
\includegraphics[width=\columnwidth]{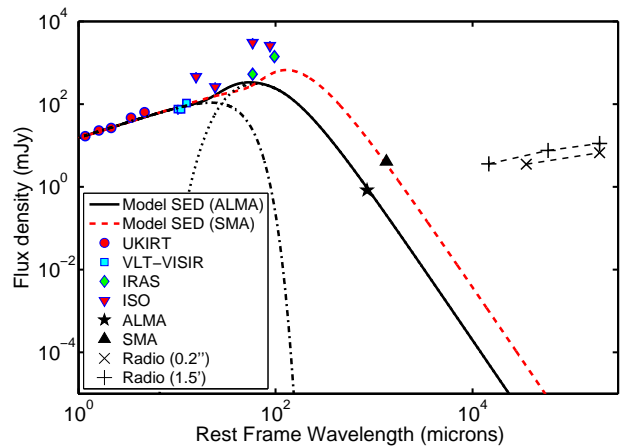}
\end{center}
\caption{{Near-IR to radio spectral energy distribution (SED) of Mrk\,590. The symbols show observed values of the continuum flux densities obtained from published work (Table~\ref{continuumfluxpub}) and from this study (ALMA and SMA continuum flux densities). The solid black curve shows the model fit to the observed IR SED and the ALMA flux density; the SMA data point is excluded (Section~\ref{thermaldust}). Also shown are the model component of the AGN power-law continuum (black dash-dotted curve) and the far-IR modified black body component (black dotted curve; details are in Appendix~\ref{appendix}). The alternate model fit to the IR SED plus the SMA flux density (ALMA data omitted; Section~\ref{SMAcont}) is shown as the red dashed curve.\label{sedfit}}} \end{figure}  

The dust temperature of the far-IR component is estimated from the model SED to be 53\,K, within the typical $\sim$40\,K to 70\,K range of temperatures observed for the `cold' dust components in other local Seyferts in the CfA sample \citep{garciaespinosa01}. Although these cool dust temperatures are often associated with star forming regions and starburst galaxies, we argue that the far-IR and 343\,GHz continuum emission is AGN dominated. This is based on the fact that the total far-IR luminosity derived from the model SED, $L_{\rm FIR} \sim 2 \times 10^{10} \, L_{\odot}$, yields ${\rm SFR}(L_{\rm FIR}) \sim 3.4 \, M_{\odot}\, \rm yr^{-1}$, four orders of magnitude higher than SFR(H$_2$) (Table~\ref{resultssummary}). Furthermore, we detect only the continuum emission in the centre with no stellar-heated dust continuum emission tracing the $^{12}$CO(3--2) gas ring.

Furthermore, there are hints that the mid-IR and far-IR continuum fluxes may exhibit variability with trends consistent with that observed in the optical-UV line and continuum emission (described in Appendix~\ref{appendix}). If the thermal dust emission is predominantly AGN heated, the variability of the mid-to-far-IR continuum fluxes (if real) can be caused by the heating and cooling of the dust surrounding the accretion disk and broad-line region \citep[e.g., from the dusty `torus',][]{urrypadovani95}. This variability may even extend into the Rayleigh-Jeans regime at mm wavelengths. Continued monitoring of the (sub-)mm, far-IR and mid-IR continuum fluxes will be crucial to confirm this short-timescale variability. Detecting variability on timescales of years to decades will also confirm a small spatial extent of the far-IR emitting regions. If the flux variations follow trends similar to the variability of the AGN observed at other wavelengths (e.g., broad emission line, optical-UV and X-ray fluxes), it provides further evidence that the mm and far-IR continuum emission is dominated by the AGN rather than stellar components at these spatial scales. 

The model SED yields a dust mass of $M_{\rm dust}\sim 10^5\, M_{\odot}$. Adopting a dust to total gas mass ratio of $\sim 0.01$ and assuming solar metallicities \citep{draineetal07}, we infer a total gas mass (both atomic and molecular) of $M_{\rm gas} \sim 10^7\,M_{\odot}$. This is more than an order of magnitude higher than the upper limits of $M({\rm H_2}) \lesssim 10^5\,M_{\odot}$ or $M({\rm H_2}) \lesssim 5 \times 10^6\,M_{\odot}$ estimated in Section~\ref{gasmass}, depending on the conversion factor, $\alpha_{\rm CO}$, used. If the actual H$_2$ to HI gas mass ratio in the central 150\,pc is equivalent to 0.48 as measured by single-dish observations \citep{maiolinoetal97}, we can explain the discrepancy to within the measurement and CO conversion uncertainties. With the HI gas mass twice the value of $M({\rm H_2})$, the total HI and H$_2$ gas mass can be as high as $\sim 5 \times 10^7\,M_{\odot}$, consistent with the value of $M_{\rm gas}$ derived from the dust mass. However, since the gas is expected to be predominantly molecular in the central regions (Section~\ref{SFR}), the adopted H$_2$ to HI gas mass ratio is unrealistic. While the value of $M_{\rm gas}$ derived from $M_{\rm dust}$ is consistent with the higher end of $M({\rm H_2})$ constraints determined from the $^{12}$CO(3--2) emission, another explanation for the possible overestimation of the gas mass from the continuum emission is that the total ALMA continuum flux may be biased by other contributing sources of emission, e.g., synchrotron emission from the AGN jet or free-free emission.

\subsubsection{Synchrotron and Free-free Emission}\label{freefreesynch}

We now examine if other sources of continuum emission besides dust, namely synchrotron and/or free-free emission, contribute significantly to the total 344\,GHz continuum flux density measured by ALMA. From the 8.4\,GHz and 344\,GHz integrated flux densities, each obtained at comparable angular resolutions of $\sim 0\farcs2$, we estimate the spectral index to be $\alpha^{8.4\,{\rm GHz}}_{344\,{\rm GHz}} \sim −0.36$. This is comparable to the spectral index of $\rm \alpha^{1.4\,GHz}_{8.4\,GHz} \sim −0.32$ obtained by \citet{theanetal01} from radio continuum observations. The radio spectral index, $\rm \alpha^{1.4\,GHz}_{8.4\,GHz}$, could arise from synchrotron emission that is self-absorbed at lower frequencies, or a combination of synchrotron (typically $\alpha \sim -0.7$) and free-free emission (typically $\alpha \sim -0.1$). These emission components could extend into the mm wavelength regime and contribute to a fraction of the ALMA continuum flux. However, \cite{edelson87} also measured a slightly steeper spectral index from the 5\,GHz and 20\,GHz flux density, $\alpha^{5\,{\rm GHz}}_{20\,{\rm GHz}} \sim -0.54$. This could mean that the radio emission is self-absorbed synchrotron emission that is beginning to steepen at around 20\,GHz. If the spectral index continues to steepen further at frequencies above 20\,GHz, it would imply that free-free emission and/or synchrotron emission are not the sole contributors to the 344\,GHz continuum emission. We have secured more observing time with the Jansky Very Large Array and ALMA in Cycle 3 to better characterize the continuum spectral shapes at (sub-)mm to radio wavelengths, and better determine the source of this continuum emission.

\subsection{Origin of SMA Continuum Emission}\label{SMAcont}

The larger synthesized beam of the SMA measures additional extended, low surface brightness continuum emission from the inner gas ring, resulting in a higher flux density at 219\,GHz compared to the 344\,GHz ALMA continuum flux (Figure~\ref{sedfit}). This emission may have been resolved out or missed due to the low surface brightness sensitivity of the ALMA images. We carry out the SED model-fitting procedure as described in Section~\ref{thermaldust} and Appendix~\ref{appendix} again, this time using the 219\,GHz SMA flux density and excluding the ALMA measurement. The model fit (red dashed line in Figure~\ref{sedfit}) gives a colder dust temperature of 22\,K and a higher dust mass of $4.4 \times 10^6 \, M_{\odot}$ (relative to the ALMA-based estimate of the dust mass in Section~\ref{thermaldust}), consistent with the scenario where the SMA continuum flux contains contributions from colder, stellar-heated dust emission. From the dust mass, we estimate a total gas mass of $M_{\rm gas} \sim 4.4 \times 10^8 \, M_{\odot}$, again assuming a dust to gas mass ratio of $\sim 0.01$ and solar metallicities \citep{draineetal07}. This gas mass is consistent with the molecular gas mass of $M({\rm H_2}) \sim 10^7 \, M_{\odot}$ to $10^8 \, M_{\odot}$ (depending on $\alpha_{\rm CO}$) estimated for the inner 4$''$ of Mrk\,590 from the CO data. However, the SED-inferred $L_{\rm FIR}$ of $2.4 \times 10^{10} \, L_{\odot}$ still overestimates the SFR, indicating that there is still a significant AGN contribution to the total 219\,GHz flux. From the $L_{\rm FIR}$, we obtain a SFR estimate of ${\rm SFR}(L_{\rm FIR}) \sim 4.1 \,\rm M_{\odot}\, yr^{-1}$, at least an order of magnitude higher than the limit of ${\rm SFR(H_2)} \lesssim 0.004 \,\rm M_{\odot}\, yr^{-1}$ to $0.2 \,\rm M_{\odot}\, yr^{-1}$ (depending on $\alpha_{\rm CO}$) derived for the inner 4$''$ in Section~\ref{SFRCO}. These additional contributions could originate from AGN-heated dust in the central 150\,pc, or from synchrotron and free-free emission arising from both the AGN and the star-forming regions in the gas ring itself. 

\subsection{Implications for Studies of AGN and Galaxy Evolution}\label{unification}

Changing-look AGNs demonstrate that it is important to account for a variable AGN accretion rate and evolution when unifying the wide range of observed AGN characteristics at all wavelengths. Some studies attempt to construct a coherent evolutionary sequence between Type\,1 AGNs (displaying broad emission lines in the spectra) and Type\,2 AGNs (with no broad emission lines) in the context of galaxy evolution. For example, \citet{villarroelkorn14} find significant differences in the properties (e.g., colors and nuclear activity) of neighbouring companions of Type\,1 and Type\,2 AGN hosts. They suggest that a significant fraction of Type\,2 AGNs are produced as a result of a merger, where dense gas and dust obscure the central accretion disk and broad-line region. The AGN then transforms into a Type\,1 once it blows away a significant fraction of the dusty torus to reveal the broad-line emission and continuum emission from the accretion disk. The results of \citet{maiolinoetal97} seem to support such a scenario, where Seyfert\,2 galaxies appear to have more disturbed CO morphologies, and higher star formation rates \citep{maiolinoetal95} relative to Seyfert\,1 galaxies. In Mrk\,590 and the quasar SDSS J015957.64+003310.5, the `type-transitioning' is reversed, demonstrating that AGN evolution can go both ways. A Type\,1 AGN can turn off when it runs out of fuel, and thus lose the broad-line emission due to a lack of ionizing photons from the accretion disk, and thus transition into a Type\,2 AGN \citep[as proposed by][]{elitzuretal14}.

The fact that such type-transitioning behavior can occur on very short timescales of years to decades, as observed in both Mrk\,590 and NGC\,4151, shows that caution should be exercised when interpreting statistical studies involving comparisons of the host galaxy properties of Type\,1 and Type\,2 AGN samples. A sample of Type\,2 AGNs selected based solely on the absence of broad emission lines, may contain a mix of nuclei that are truly obscured, as well as unobscured nuclei that may be undergoing low accretion states or that may have turned off. The significance of such biases in studies of AGN and galaxy evolution will depend on how prevalent changing-look AGNs are, how often such type-transitioning behavior occurs, and their duty cycles (i.e., the relative timescales at which an AGN nucleus exhibits Type\,1 properties compared to that of a Type\,2). These quantities are unknown at present.

The OzDES reverberation mapping program \citep{kingetal15} will monitor the broad emission lines of $\sim 500$ quasars over a five year period. Such long-term spectroscopic surveys may potentially uncover more of these changing-look AGNs. Follow-up imaging of the CO line emission of such changing-look AGNs with ALMA will allow their gas reservoirs and fueling processes to be studied and compared with that of the general population of active galaxies on a statistical level. Such studies will determine how unique these changing-look AGNs are, and promise to provide a better understanding of the intermittency of nuclear activity as well as the physics of black hole fueling in general.

\section{Summary and Conclusions}\label{conclusion}
   
In this study, we investigate if the AGN in Mrk\,590 is turning off due to a lack of central gas to fuel it. We present the first interferometic maps of the molecular gas distribution (traced by the $^{12}$CO(3--2) emission line) and (sub-)mm continuum emission in the circumnuclear regions of Mrk\,590. The $^{12}$CO(3--2) and 344\,GHz continuum maps at sub-arcsecond resolution were obtained with ALMA, while the 219\,GHz continuum map at 3$''$ resolution was obtained with the SMA. The main results of our study are summarized as follows:

\begin{enumerate}
 
\item We do not detect any $^{12}$CO(3--2) line emission in the central 150\,pc down to the sensitivity limit of the observations. This constrains the molecular gas mass in the centre to $M({\rm H_2}) \lesssim 1.6 \times 10^5\, {M_{\odot}}$ assuming $\alpha_{\rm CO}\sim 0.8\,{M_{\odot}\,\rm (K \,km\,s^{-1}\,pc^{2}})^{-1}$ as observed in the centres of nearby galaxies, and $L'_{\rm CO(3-2)}/L'_{\rm CO(1-0)} \sim 0.9$ as observed in nearby Seyfert galaxies. Higher sensitivity observations with the full ALMA array will provide even stronger constraints on the gas mass. Future observations and detections of other CO transitions will significantly reduce the uncertainties in the H$_2$ mass estimates arising from presently unknown excitation levels of the CO gas. 

\item Although we find relatively low levels of $M({\rm H_2})$ and ${\rm SFR(H_2)} \lesssim 2 \times 10^{-4}\,M_{\odot}\, \rm yr^{-1}$ (estimated based on $M({\rm H_2})$ and using the Kennicutt-Schmidt relation) in the central 150\,pc, there is still potentially enough gas to fuel the AGN for a further $2.6 \times 10^5$ years even if it accretes at the Eddington limit. We therefore cannot rule out the possibility that Mrk\,590 may turn on again in the near future.  

\item We observe a ring-like structure of molecular gas at a radius of 1\,kpc, tracing regions containing faint dust lanes seen in \textit{HST}/ACS F550M (\textit{V}-band) images. We measure a molecular gas mass of order ${M({\rm H_2})} \sim 10^7 \,M_{\odot}$ in the central 2\,kpc, comparable to that observed in other nearby Seyfert galaxies at similar scales. Mrk\,590 therefore does not have significantly less molecular gas at these scales, confirming that large scale structures and gas reservoirs provide no direct indication of ongoing AGN activity and accretion rates.

\item A molecular gas component with disturbed kinematics centred just $\sim 200$\,pc west of the AGN may be fueling the centre. This component has sufficient H$_2$ gas to fuel the central engine for up to $10^6$ years at the Eddington-limited rate. Spatially resolved gas kinematics from higher resolution CO maps from ALMA, and from more emission lines, e.g., $\rm \left[ O\,{\textsc{iii}} \right]$ and HI, at tens of pc to kpc scales, are required to better model the galaxy rotation curves and central gas dynamics. These additional data, which we are in the process of obtaining, will enable us to better determine the gas transport mechanisms potentially operating at these spatial scales in this source.  

\item We do not detect any strong molecular outflows (with velocities greater than the rotational velocities) that may be depleting the gas in the centre, down to the detection limits of our $^{12}$CO(3--2) spectra.

\item The HCO$^+$(4--3) line is not detected in the ALMA observations. We obtain a HCO$^+$ to CO integrated flux ratio of $R^{\rm HCO43}_{\rm CO32} \lesssim 0.14$ in the inner 2\,kpc. This limit on the dense gas fraction in Mrk\,590 is consistent with that found in other nearby AGN host galaxies. The $^{12}$CO(2--1) emission line was also not detected in the SMA spectra, consistent with $^{12}$CO(3--2) to $^{12}$CO(2--1) flux ratios of $\gtrsim 2$ as observed in quasar host galaxies, sub-mm galaxies, and the Milky Way Galaxy.

\item We detect marginally resolved continuum emission at 344\,GHz in the centre that is spatially coincident with the unresolved cm-wavelength radio source, where the AGN is likely located. The origin of the emission is most likely thermal dust emission ($T \sim 50$\,K) that is heated predominantly by the AGN, possibly from the dusty torus posited in AGN unification models. However, contributions from free-free and synchrotron emission cannot be ruled out. Since the emission is AGN-dominated, its flux density is likely to be variable on timescales of years or decades, as has been observed in the mid-to-near-IR, optical-UV and X-ray wavelengths. Continued monitoring of the (sub-)mm flux will confirm this, and provide a better understanding of the energy coupling of the various AGN components emitting at various wavelengths. 
  
 \end{enumerate}

\section*{Acknowledgements}

We are extremely grateful to Lyuba Slavcheva-Mihova for providing us with their \textit{HST} structure map of Mrk\,590. Also, we thank Sandra Raimundo for helpful discussions and the anonymous reviewer for insightful comments. We wish to thank the staff of the Nordic and Italian nodes of the European ALMA Regional Centre (ARC), in particular Ivan Marti-Vidal, for assistance with the data reduction and imaging. We are also indebted to SMA support staff, including Mark Gurwell and Lars Kristensen, for their assistance with the preparation of the SMA proposal, scheduling of observations and data reduction. Travel to the Nordic node of the ARC was made possible through the MARCUs RadioNet3 funding from the European Commission Seventh Framework Programme (FP/2007-2013) under grant agreement No. 283393. JYK is supported by a research grant (VKR023371) from Villumfonden. MV, JYK, and DL gratefully acknowledge support from the Danish Council for Independent Research via grant no. DFF 4002-00275. The Dark Cosmology Centre is funded by the Danish National Research Foundation. The work of VC is partly supported by the Italian Ministero dell\'\,Istruzione, Universit\`a e Ricerca through the grant Progetti Premiali 2012 -- iALMA (CUP C52I13000140001). BMP is grateful to the US National Science Foundation for support through grant AST--1008882 to The Ohio State University. This paper makes use of the following ALMA data: ADS/JAO.ALMA\#2013.1.00534.S. ALMA is a partnership of ESO (representing its member states), NSF (USA) and NINS (Japan), together with NRC (Canada), NSC and ASIAA (Taiwan), and KASI (Republic of Korea), in cooperation with the Republic of Chile. The Joint ALMA Observatory is operated by ESO, AUI/NRAO and NAOJ. The National Radio Astronomy Observatory (NRAO) is a facility of the National Science Foundation operated under cooperative agreement by Associated Universities, Inc. This research also makes use of data from the SMA, which is a joint project between the Smithsonian Astrophysical Observatory and the Academia Sinica Institute of Astronomy and Astrophysics and is funded by the Smithsonian Institution and the Academia Sinica. We also used data based on observations made with the NASA/ESA Hubble Space Telescope, and obtained from the Hubble Legacy Archive, which is a collaboration between the Space Telescope Science Institute (STScI/NASA), the Space Telescope European Coordinating Facility (ST-ECF/ESA) and the Canadian Astronomy Data Centre (CADC/NRC/CSA). The NVAS image used in this paper was produced as part of the NRAO VLA Archive Survey, (c) AUI/NRAO. We acknowledge the use of data from the NASA Extragalactic Database (NED).

%%%%%%%%%%%%%%%%%%%%%%%%%%%%%%%%%%%%%%%%%%%%%%%%%%



%\bibliographystyle{mnras}


%%%%%%%%%%%%%%%%%%%%%%%%%%%%%%%%%%%%%%%%%%%%%%%%%%

%%%%%%%%%%%%%%%%% APPENDICES %%%%%%%%%%%%%%%%%%%%%

\appendix

\section{IR Continuum SED Modelling and Caveats}\label{appendix}

We describe here the details of our model fit to the observed near-IR to (sub-)mm continuum fluxes from published data (Table~\ref{continuumfluxpub}) and our ALMA observations, as discussed in Section~\ref{thermaldust}. The model fitting is based on the far-IR SED fitting code developed by \citet{casey12}. The code combines a far-IR greybody spectrum of a single temperature, with a near-IR to mid-IR power law function to approximate warm dust emission due to AGN heating. We fix the mean emissivity index of dust particles at the typical value of $\beta = 1.5$, since we are unable to constrain it well through the model-fitting process itself with only a single point at wavelengths of $\lambda \gtrsim 200\,\mu$m (from ALMA). This value of $\beta$ has been confirmed by observations of our Galaxy \citep{dupacetal03} and nearby starburst galaxies \citep{kovacsetal10}, and is consistent with values measured in the laboratory \citep{agladzeetal96}. The other five parameters, including the temperature of the greybody component and the slope of the mid-IR power law, are kept as free parameters during the model-fitting process. The fitting procedure, the model SED and the model parameters are described in further detail by \citet{casey12}.

We note that there will be large uncertainties associated with the SED model-fitting, and in the parameters (e.g., dust temperatures, dust masses and total far-IR luminosity) derived from the fitted model. These uncertainties arise from:
\begin{enumerate}
\item \textit{Inconsistent beam sizes and angular resolution.} The UKIRT, \textit{ISO} and \textit{IRAS} flux densities are measured using larger beam sizes compared to our ALMA measurements. Even though the near-IR to far-IR flux densities from these instruments are very likely AGN dominated, contamination from emission originating at kpc to tens of kpc scales cannot be ruled out. 

\item \textit{Systematic errors in the absolute flux scaling of the different instruments.} Since the SED model fit makes use of a heterogenous dataset obtained from many different instruments, there are bound to be indeterminable uncertainties in the absolute scaling of the flux measurements from each telescope. For example, there are known discrepancies between \textit{ISO} flux measurements at 90\,$\mu$m and \textit{IRAS} flux measurements at 100\,$\mu$m. \citet{serjeanthatziminaoglou09} find evidence that \textit{IRAS} fluxes at 100\,$\mu$m can be overestimated by up to $\sim$30\% compared to \textit{ISO} and \textit{Spitzer} fluxes at comparable wavelengths.  

\item \textit{Possible intrinsic variability of the AGN continuum emission.} There are indications that the far and mid-IR continuum flux densities may exhibit variability consistent with the changes in the optical-UV continuum and line flux densities as reported by \citet{denneyetal14}. The \textit{ISO} observations in the 1990s \citep{garciaespinosa01} show that the 60\,$\mu$m and 90\,$\mu$m flux densities have increased by factors of a few when compared to the \textit{IRAS} observations in the 1980s \citep{edelsonetal87}. This is consistent with the flux density increase observed at other wavelengths during this period. This is despite the smaller beam size of \textit{ISO} which measures the flux density at smaller angular scales relative to \textit{IRAS}. As mentioned above, \textit{ISO} flux density measurements are also typically found to be consistently lower than that of \textit{IRAS}. Therefore, this systematic error cannot explain the higher \textit{ISO} flux densities compared to the \textit{IRAS} flux densities measured for Mrk\,590. Furthermore, the AGN-dominated 12.8\,$\mu$m flux densities measured in 2006 with VLT-VISIR \citep{horstetal09} are a factor of four lower than the 16\,$\mu$m \textit{ISO} flux densities measured in the 1990s, consistent with the flux density decrease observed at other wavelengths over this time period. While angular resolution effects cannot be ruled out in this case, we note that variability may also have affected the VLT-VISIR data. 

\end{enumerate} 

While the far-IR flux densities of the model fit are consistently lower than the observed \textit{ISO} and \textit{IRAS} flux densities (Figure~\ref{sedfit}), the larger beam sizes of these instruments (relative to that of ALMA and VLT-VISIR) imply that their measured flux densities are only upper limits of the flux emanating from within the central 150\,pc. Although UKIRT also has a lower angular resolution of $\sim 5''$, the emission at wavelengths from 1\,$\mu$m to 10\,$\mu$m is expected to be completely dominated by the central AGN. Another possible explanation for the higher ISO fluxes is the possible variability of the far-IR continuum emission. In fact, far-IR continuum variability provides a better explanation (as opposed to angular resolution effects) for the higher \textit{ISO} fluxes (relative to the model fit values) as seen in Figure~\ref{sedfit}. With the exception of the $0\farcs1$ radio data (crosses in Figure~\ref{sedfit}), the \textit{ISO} fluxes are the only measurements obtained in the 1990s when Mrk\,590 was accreting at its highest observed rate. All the other observational data points were obtained either in the early 1980s or from the year 2000 onwards when Mrk\,590 was accreting at a lower rate. We therefore argue that the model SED is a reasonable fit to the data.

%%%%%%%%%%%%%%%%%%%%%%%%%%%%%%%%%%%%%%%%%%%%%%%%%%

% Don't change these lines
\bsp	% typesetting comment
\label{lastpage}
\end{document}